\documentclass[12pt]{article}

\usepackage{scicite}
\usepackage{graphicx}

\usepackage{times}

\newenvironment{sciabstract}{%
\begin{quote} \bf}
{\end{quote}}

\title{Google Scholar is manipulatable} 
\author
{Hazem Ibrahim,$^{1, 2,\ddagger}$ Fengyuan Liu,$^{1, 3, \ddagger}$ Yasir Zaki$^{1,\ast}$  Talal Rahwan$^{1,\ast}$ \\
\\
\normalsize{$^{1}$Department of Computer Science, New York University Abu Dhabi, UAE}\\
\normalsize{$^{2}$Tandon School of Engineering, New York University, New York, USA}\\
\normalsize{$^{3}$Courant Institute of Mathematical Sciences, New York University, New York, NY 10012, USA}\\
\normalsize{$^\ddagger$ Joint first authors
}\\
\normalsize{$^\ast$To whom correspondence should be addressed; E-mails: \{talal.rahwan, yasir.zaki\}@nyu.edu}
}

\date{}

\begin{document} 

\baselineskip24pt

\maketitle 

\begin{sciabstract}
Citations are widely considered in scientists' evaluation. As such, scientists may be incentivized to inflate their citation counts. While previous literature has examined self-citations and citation cartels, it remains unclear whether scientists can purchase citations. Here, we compile a dataset of $\sim$1.6 million profiles on Google Scholar to examine instances of citation fraud on the platform.
We survey faculty at highly-ranked universities, and confirm that Google Scholar is widely used when evaluating scientists. Intrigued by a citation-boosting service that we unravelled during our investigation, we contacted the service while undercover as a fictional author, and managed to purchase 50 citations. These findings provide conclusive evidence that citations can be bought in bulk, and highlight the need to look beyond citation counts.


\end{sciabstract} 


\newpage
\section*{Significance statement}
In this study, we analyze over 1.6 million Google Scholar profiles and identify instances of citation manipulation to boost an author's citation metrics. In many cases, this was done through a previously undocumented practice of ``citation purchasing'', where an author pays a small fee to a third party company which provides citations to the author in bulk. We confirm that this practice is possible by purchasing citations to a fictional author. This type of manipulation is of particular concern due to Google Scholar's wide-spread use in scientist evaluation processes, which we confirm with a survey of faculty from top-ranked universities. Our findings bring to light new forms of citation manipulation and emphasize the need to look beyond citation counts. 

\section*{Introduction}
Due to the nature of scientific progress, scientists build on, and cite, the work of their predecessors to acknowledge their intellectual contributions while advancing the frontiers of human knowledge~\cite{wang2021science}.
However, the purpose of citations has evolved beyond this initial purpose, with it arguably becoming a synonym for scientific success~\cite{acuna2012predicting,sinatra2016quantifying}. To date, citation-based measures (e.g., h-index, i10-index, journal impact factors) dominate the evaluation of papers~\cite{crew2019google}, scientists~\cite{stephan2017reviewers} and journals~\cite{garfield1972citation}, despite the ongoing debate surrounding their use~\cite{DORA,hicks2015bibliometrics,sugimoto2021scientific}.

One reason behind using these measures is the fierce competition that scientists face when competing for limited resources such as grants, laboratory space, tenure, and the brightest graduate students~\cite{maddox1993competition}, the allocation of which requires a measure of performance~\cite{vzelezny2023competition}. In this process, bibliometric data such as citation counts have become an indispensable component of scientific evaluation for at least the past four decades~\cite{stephan2017reviewers}, and has become one of the primary metrics which scientists strive to maximize~\cite{fire2019over}.
While some may consider such a competitive environment to be a force for good~\cite{franck1999scientific}, others argue that it creates perverse incentives for scientists to ``game the system'', with many resorting to various methods to boost their citation counts, e.g., via self-citations~\cite{hyland2003self,ioannidis2015generalized}, citation cartels~\cite{fister2016toward,perez2019network,kojaku2021detecting,oransky2017gaming}, and coercive citations~\cite{wilhite2012coercive,herteliu2017quantitative,thombs2015potentially,martin2013whither}. Together, these practices are known as citation manipulation or reference list manipulation. Unfortunately, these practices are hard to define due the fact that the line between innocuous and malicious practices is not always clear. For example, every scientist builds upon their own research or the work of their collaborators, so everyone is expected to self-cite or frequently exchange citations with their collaborators. As a result, it is virtually impossible to determine with absolute certainty whether someone cites out of necessity or simply to game the system.

Perhaps the most egregious approach to manipulating one's citations would be to purchase them outright. Anecdotes of emails asking authors to cite articles in return for a fee have been exposed~\cite{adam2021publisher,Myriam2022new}, although it is unclear whether these emails were simply scams. As such, it remains unclear whether it is possible for any given scientist to purchase citations to their own work. For such a transaction to take place, it requires at least two culprits that are research-active scientists---the one who purchases the citations, and the one who plants them in their own articles in return for a fee; a possible third culprit would be an individual or company who brokers this transaction. 

Against this background, we set out to understand the proportion of faculty members in highly-ranked universities who use citation metrics when recruiting or promoting candidates, and the databases they use to retrieve these metrics. To do so, we surveyed over 30,000 faculty members of the 10 highest ranked universities globally. We found that the majority of faculty members indeed consider citation counts, and Google Scholar is the most popular source to retrieve such information, beating all other databases combined. Surprisingly, despite the popularity of Google Scholar, it has received little attention in studies investigating scientists and their citations, compared to other databases such as
Web of Science~\cite{wang2013quantifying,sekara2018chaperone,nielsen2017one,wuchty2007increasing,huang2020historical,wu2019large,fleming2019government,ahmadpoor2017dual}, 
USPTO~\cite{wu2019large,fleming2019government,ahmadpoor2017dual},
Microsoft Academic Graph~\cite{huang2020historical,li2022untangling,alshebli2022beijing,kojaku2021detecting,alshebli2018preeminence,liu2023gender},
or discipline-specific databases~\cite{huang2020historical,yin2019quantifying,wang2013quantifying,liu2023non,petersen2015quantifying,jia2017quantifying,azoulay2010superstar,nielsen2017one},
possibly due to the lack of an accessible dataset which compiles all information on Google Scholar. Motivated by this finding, we steered our analysis of ``citation boosting'' towards Google Scholar and curated a dataset of over 1.6 million profiles on this platform. Using various measures of citation irregularity, we found several scientists with suspicious profiles that may have engaged with citation boosting services. Following this trail, we indeed found such a company that provided citations to one of the suspicious scientists. Before engaging with this company, we generated 20 research articles using a Large Language Model while listing a fictional character as an author. We then created a Google Scholar profile for that character, and contacted the company to ask them to boost the citations of this profile. Indeed, we managed to purchase 50 citations, proving that citations can indeed be bought in bulk for a relatively small fee in a matter of weeks. Together, our findings question the reliance on Google Scholar to evaluate scientists by demonstrating its susceptibility to manipulation, while contributing more broadly to the plea of reducing the reliance on citation metrics when evaluating scientists.

\section*{Results}
\subsection*{Google Scholar is a prominent source of citation metrics}
\begin{figure}[!htbp]
    \centering
    \includegraphics[width=\textwidth]{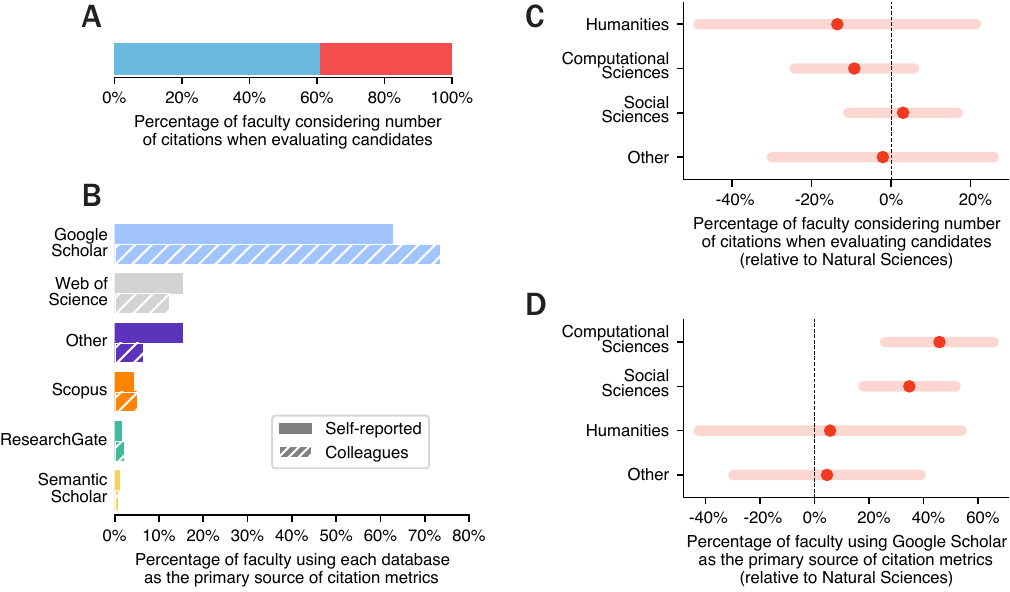}
    \caption{
\textbf{Survey responses from faculty of the top-10 ranked universities around the world.} 
\textbf{A}, The percentage of faculty who consider citations when evaluating candidates (blue) and those who do not (red). 
\textbf{B}, Solid bars indicate, out of those who self-report considering citations when evaluating candidates, the percentage of faculty using each database as the primary source of citation metrics. Hatched bars indicate, out of those who report that their colleagues consider citations when evaluating candidates, the percentage of colleagues using each database.
\textbf{C}, Relative to the Natural Sciences, the percentage of faculty from each group of disciplines who consider citations when evaluating candidates. 
\textbf{D}, Relative to the Natural Sciences, the percentage of faculty who use Google Scholar as the primary source of citation metrics.
In (\textbf{C}) and (\textbf{D}), dots denote OLS-estimated coefficients and error bars represent 95\% confidence intervals. 
    }
    \label{fig:survey}
\end{figure}

To gauge the extent to which academics consider citations when evaluating scientists for recruitment or promotion purposes, we surveyed faculty members of the 10 most highly ranked universities around the world (see Methods for more details). The results of this survey are summarized in Figure~\ref{fig:survey}. We find that the majority of faculty consider the number of citations when evaluating candidates (Figure~\ref{fig:survey}A). Moreover, out of those who consider citations, over 60\% obtain this data from Google Scholar, making it more popular than all other sources combined (Figure~\ref{fig:survey}B, solid bars). Participants were also asked about the source of citation metrics they believe their colleagues use the most. Again, Google Scholar is the most popular source, selected by 73\% of the participants (Figure~\ref{fig:survey}B, hatched bars).

Next, we investigate whether the usage of citation metrics and Google Scholar differ across disciplines. To this end, we asked participants to select their discipline from a pre-determined list, or select ``Other'' if they did not identify with any. We then grouped the disciplines into four categories, namely, Natural Sciences, Social Sciences, Computational Sciences, and Humanities. As can be seen in Figure~\ref{fig:survey}C, we do not find evidence that disciplines differ in their likelihood to consider citations when evaluating candidates. However, when it comes to the usage of Google Scholar, we find significant differences across disciplines (Figure~\ref{fig:survey}D). Faculty in the Computational Sciences and Social Sciences are more likely to use Google Scholar as their primary source of citation metrics, compared to those in the Natural Sciences ($p < 0.001$).

Overall, our survey indicates that citations are often used by faculty members when evaluating scientists for recruitment or promotion purposes. Moreover, it establishes for the first time that Google Scholar is by far the most popular source of citation metrics used for this purpose.

\subsection*{Investigating suspicious citations}
Our findings thus far suggest that scientists are incentivized to increase their citations on Google Scholar. Next, we investigate whether scientists can manipulate their Google Scholar profiles by planting fake citations. Here, we follow a proof-by-example approach. That is, our goal is to identify a few examples of Google Scholar profiles that seem highly suspicious, thereby demonstrating that the platform can indeed be manipulated by scientists without having their profiles suspended. To this end, we identified five suspicious authors whose Google Scholar profiles exhibit highly irregular patterns, and compared them against authors who are similar in terms of their field of research, academic birth year, number of publications, and number of citations; see Methods for more details on how the suspicious authors and their matches are identified.

\begin{figure}[!htbp]
    \centering
    \includegraphics[width=\textwidth]{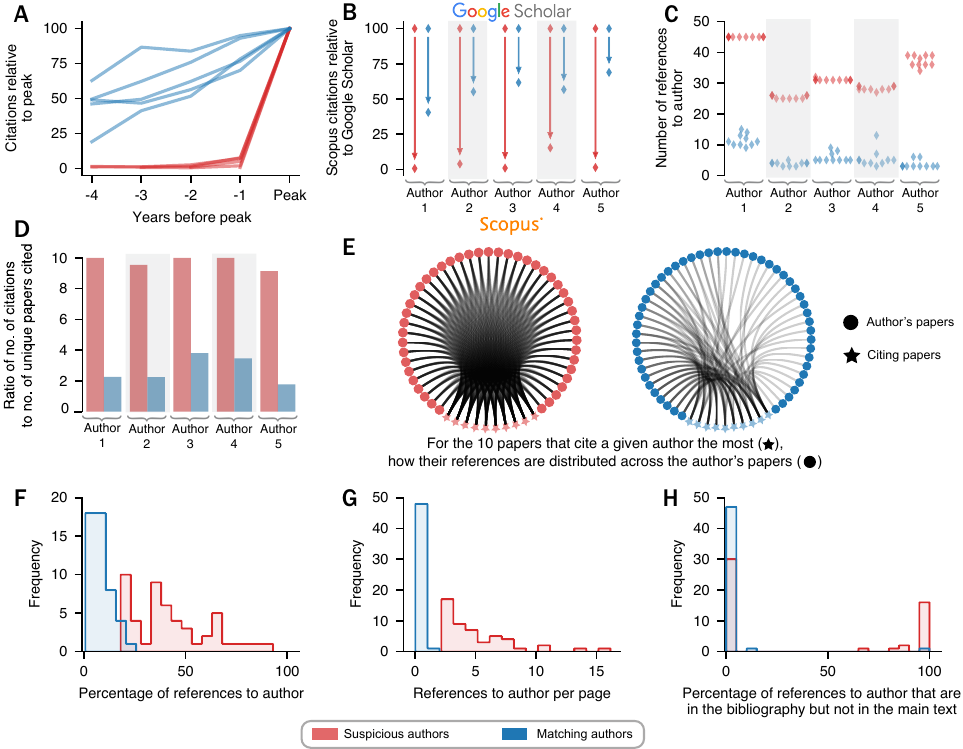}
    \caption{\textbf{A comparative analysis of suspicious authors and their matches.} In each plot, red lines and red dots denote suspicious authors, while blue ones denote their matches.
    \textbf{A}, For the 4 years leading up to an author's peak citations, the annual number of citations relative to the peak.
    \textbf{B}, Discrepancy between Google Scholar and Scopus in terms of the author's citation count in their peak year.
    In (\textbf{C})-(\textbf{H}), for any given author, we focus on their 10 ``citing papers'', i.e., the ones that reference them the most.
    \textbf{C}, The number of references to a given author in each of their 10 citing papers.
    \textbf{D}, The total number of references pointing to an author from their 10 citing papers, divided by the number of the author's unique papers being cited therein.
    \textbf{E}, The citation network between the 10 citing papers ($\star$) and an author's papers ($\circ$) for a suspicious author (red) and their matching author (blue).
    \textbf{F}, For each value $v$ on the x-axis, how many of the citing papers have $v\%$ of their references pointing to the author in question.
    \textbf{G}, For each value $v$ on the x-axis, how many of the citing papers have an average of $v$ references per page pointing to the author in question.
    \textbf{H}, For each value $v$ on the x-axis, how many of the citing papers have $v\%$ of their references pointing to the author in question and are not referenced in the citing papers' main text despite being listed in its bibliography.
}
\label{fig:sus_authors}
\end{figure}

The results of the comparison between suspicious authors (red) and their matches (blue) are summarized in Figure~\ref{fig:sus_authors}. We begin our analysis by comparing their annual citation rates prior to the year in which they reached their peak number of citations (Figure~\ref{fig:sus_authors}A). Here, to unify the scale across the authors, we only depict their number of citations relative to their peak (see Supplementary Figure~1 for the absolute numbers). As can be seen, the matched authors exhibit a gradual increase in citations prior to their peak, whereas the suspicious authors exhibit a sudden spike in their citations in their peak year.

Next, we compare the authors' citation counts in their peak year as listed in Google Scholar vs.\ Scopus (Figure~\ref{fig:sus_authors}B), since the latter only indexes journals that are reviewed and selected by an independent Content Selection and Advisory Board~\cite{scopus}. While it is natural to see a drop in citation counts due to the more restrictive indexing by Scopus, suspicious authors on average experience a citation drop of 96\% on Scopus, while normal authors on average experience a drop of 43\%.
In other words, not only do the suspicious authors experience an unexpected increase in their citation counts in a particular year, but more than 96\% of those citations cannot be found on Scopus.

Our next analysis focuses on the 10 papers that contain the highest number of citations to each author, hereinafter referred to as the ``citing papers''. As can be seen in Figure~\ref{fig:sus_authors}C, suspicious authors are referenced up to 45 times in a citing paper, while their matches are referenced no more than 15 times. Additionally, the variance in the number of times a suspicious author is referenced by citing papers is suspiciously small. For example, the first suspicious author is referenced exactly 45 times by all their citing papers. 
It would be even more suspicious if all citing papers are referencing the same 45 papers of the author, rather than each of them citing 45 different papers of that author.
To determine whether this is the case, Figure~\ref{fig:sus_authors}D depicts the total number of references pointing to an author from their 10 citing papers, divided by the number of the author’s unique papers being cited therein. As can be seen, for suspicious authors, this ratio is very close to 10, indicating that all references are indeed directed towards the same set of papers. On the other hand, the ratio is much smaller for the matched scientists. To better illustrate this phenomenon, Figure~\ref{fig:sus_authors}E shows two bipartite graphs of the 10 citing papers (stars) and the set of unique authored papers (circles) for the suspicious author~\#1 and their match. Here, the left panel is a complete bipartite graph, indicating that all 10 citing papers are referencing the exact same set of 45 papers, while the right panel is far from complete. 

One possible explanation of the discrepancy in citations between suspicious authors and their matches in Figure~\ref{fig:sus_authors}C is that the citing papers of the suspicious authors have more pages and/or have more references, compared to those of the matched authors. However, we find that this is not the case. More specifically, the suspicious authors tend to receive much more citations even when accounting for the total number of references in the bibliography (Figure~\ref{fig:sus_authors}F) and the total number of pages (Figure~\ref{fig:sus_authors}G) in the citing papers. Also noteworthy is the fact that the percentage of the references to a suspicious author reaches as high as 90\% in one of the citing papers; see Supplementary Appendix~1 for a redacted version of that paper. 

Finally, in many citing papers, the references made to a suspicious author are nowhere to be found in the main text, but only in the bibliography. As shown in Figure~\ref{fig:sus_authors}H, in 18 of the 50 papers analyzed for the suspicious authors, all of the references made to the author do not appear in the main text, indicating that all those references made to the suspicious authors are irrelevant to the content of the citing paper, which in turn suggests that they were planted to boost the citation count of the suspicious author. For example, one citing paper consists of only two pages, with just a single reference in its main text, but references the suspicious author 29 times in its bibliography; see Supplementary Appendix~2 for a redacted version of that paper.

\begin{figure}[!htbp]
    \centering
    \includegraphics[width=\linewidth]{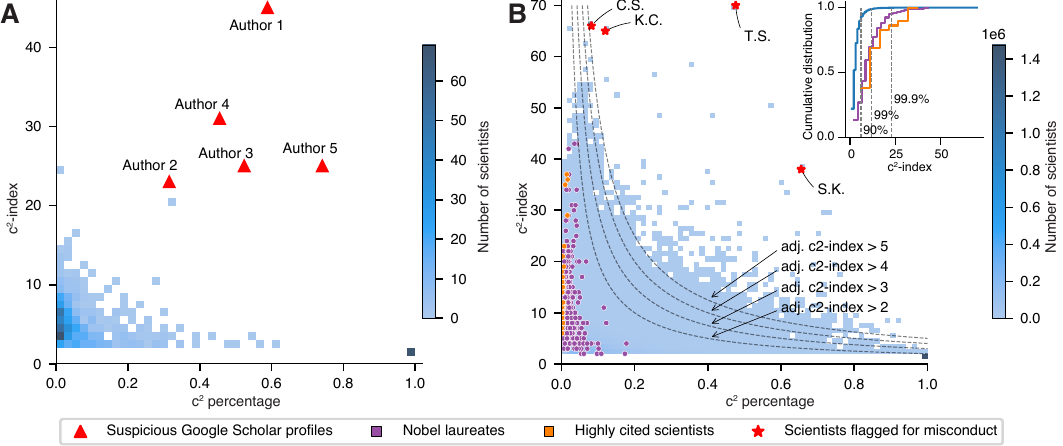}
    \caption{\textbf{The distribution of $c^2$-index.}
    \textbf{A}, A sample of 900 authors randomly selected from the nine disciplines with the largest number of authors in our dataset (100 per discipline). Triangles highlight the five suspicious authors whose Google Scholar profiles exhibit highly irregular patterns.
    \textbf{B}, Distribution of all scientists in Microsoft Academic Graph (MAG) who are cited at least 100 times. For a scientist whose $c^2$-index is $n$ ($y$-axis), the $x$-axis shows the percentage of citations a scientist receives from papers citing them at least $n$ times. 
    Inset shows the cumulative distribution of $c^2$-index of all scientists (blue), Nobel laureates (purple), and top-10 most cited scientists in the following three fields on Google Scholar (orange): machine learning, neuroscience, and bio-informatics. Red triangles highlight scientists who were previously reported to have engaged in academic misconduct.
    } 
\label{fig:c2index}
\end{figure}

Considering that suspicious authors tend to receive an excessive number of citations from a small set of papers, we propose a metric to highlight potentially suspicious profiles. In particular, we introduce the \textit{citation concentration} index, or $c^2$-index for short. Specifically, the $c^2$-index of a scientist is the largest number $n$ such that there exists $n$ papers that each cite the scientist at least $n$ times. For a scientist with a $c^2$-index of $n$, we also calculate the percentage of citations that this scientist receives from the papers that cite them at least $n$ times (henceforth, the $c^2$-percentage). To understand the distribution of scientists across these two dimensions, we randomly select 100 Google Scholar profiles from each of the nine disciplines with the largest number of authors in our dataset, and calculate their $c^2$-index and $c^2$-percentage. The results of this analysis are illustrated in Figure~\ref{fig:c2index}A. As can be seen, the vast majority of scientists fall in the bottom right corner of the distribution, since they are cited at most once by any citing paper. Very few scientists have a $c^2$-index higher than 10, with the highest $c^2$-index observed in the random sample being 25. However, among the suspicious authors, the lowest $c^2$-index is 23, while the highest is 45, i.e., there exists 45 different papers which cite this author at least 45 times each. Hence, this analysis confirms that all five suspicious authors highlighted in Figure~\ref{fig:sus_authors} have unusually high $c^2$-indices, suggesting that the $c^2$-index could help identify potentially suspicious profiles. 

So far, we only examined the $c^2$-index of a small set of authors on Google Scholar. This is due to the expensive nature of collecting citation information on Google Scholar, requiring a number of requests on the order of the total number of citations of the set of authors to retrieve the necessary information to compute the index. As such, to apply this index at scale, we resort to the Microsoft Academic Graph (MAG) dataset, which includes over 200 million scientific papers published since the year 1800. Importantly, the indices computed via MAG should be taken as a lower bound on the indices computed via Google Scholar. This is not only due to the fact that MAG has been deprecated since the end of 2021, but also due to MAG's stricter indexing policy (e.g., MAG does not include citations stemming from a document uploaded on ResearchGate or OSF).

Figure~\ref{fig:c2index}B shows the distribution of all scientists with at least 100 citations along two dimensions---their $c^2$-index and $c^2$-percentage according to MAG. As can be seen, the boundary of the distribution resembles a hyperbola---on the one hand, the vast majority of scientists are cited only once by all citing papers, while on the other hand, successful scientists (i.e., Nobel laureates or highly-cited scientists) tend to have high $c^2$-indices but low $c^2$-percentages. Notice that a scientist may have a high $c^2$-index if they produce a body of impactful works that are often cited together. However, even for these extremely successful scientists, the total number of citations that come in bulk rarely goes beyond 10\% of their total number of citations (98.6-th percentile value), and never exceeds 20\%. As a result, it is rare to have a high $c^2$ index (the 99-th percentile value of $c^2$-index is 12, as can be seen in Figure~\ref{fig:c2index}B inset), and extremely rare to have both a high $c^2$-index and a high $c^2$-percentage simultaneously. 

Meanwhile, there exists a small group of scientists who not only received citations in bulk from a large number of papers (i.e., having a high $c^2$-index), but also received a large percentage of their citations in this manner, raising the possibility that some of these citations may not be genuine. To quantify the number of scientists who have a high $c^2$-index and a high $c^2$-percentage simultaneously, we calculate an adjusted $c^2$-index by multiplying the $c^2$-index of a scientist by the $c^2$-percentage of that scientist. An example of a scientist with an adjusted $c^2$-index of 4 could be a scientist with a $c^2$-index of 40 (i.e., there are 40 papers that each cite this scientist at least 40 times) and a $c^2$-percentage of 10\% (i.e., the citations received from those 40 papers constitute one tenth of all citations accumulated by that scientist throughout their entire career). In other words, 1 in every 10 of their total citations stem from a paper that cites them 40 times or more. As for scientists with an adjusted $c^2$-index of 2, one such example could be those with a $c^2$-index of 25 and a $c^2$ percentage of 8\%. The dotted lines in Figure~\ref{fig:c2index}B show the cut-offs with varying adjusted $c^2$-indices. Across MAG, the adjusted $c^2$-index is $\geq$ 2 for 14,441 scientists, $\geq$ 3 for 3,525 scientists, $\geq$ 4 for 1,161 scientists, and $\geq$ 5 for 485 scientists. Supplementary Figure~2 depicts the distribution of adjusted $c^2$-index in MAG. As can be seen, the distribution resembles a straight line on a log-log scale, which is the signature of a power-law distribution. Finally, let us comment on Nobel laureates. As can be seen in Figure~\ref{fig:c2index}B, while some Nobel laureates have high $c^2$-indices, exceeding 40 in some cases, none of them have an adjusted $c^2$-index greater than 2.

To demonstrate that the adjusted $c^2$-index can serve as an indicator of scientists who may have engaged in questionable practices to accumulate citations, we examined four scientists who have been involved in public scandals, to determine whether they have high adjusted $c^2$-indices. More specifically, these four scientists either (1) repeatedly manipulated the peer-review process as an editor to gain citations~\cite{van2020highly}; (2) engaged in extreme self-citing practices~\cite{van2019hundreds}; (3) manipulated the peer-review process as a guest editor which resulted in the retraction of hundreds of papers in bulk~\cite{oransky2022exclusive}; or (4) engaged in extreme self-publishing behaviour by publishing more than 150 papers in a journal that they edit~\cite{pak2022innovation}. We examined the adjusted $c^2$-indices of these scientists, and found all of them to be greater than five; see the red stars in Figure~\ref{fig:c2index}B.

It is important to note that a high adjusted $c^2$-index does not necessarily signal misconduct; after all, a pioneer in their field could end up being cited tens of times per paper. Nevertheless, as we have demonstrated through real examples of public scandals, a high adjusted $c^2$-index could serve as a warning sign of citation manipulation. Next, we investigate the possible ways in which citations could be manipulated on Google Scholar.

\subsection*{Building a fictional Google Scholar profile}
We examined the 114 authors in our dataset whose Google Scholar profiles exhibit highly irregular patterns, and found that they received citations from non-peer reviewed sources more frequently than regular authors; see Supplementary Figure~4. More specifically, we find that pre-print servers, namely, arXiv~\cite{arxiv}, Authorea~\cite{authorea}, OSF~\cite{OSF}, and ResearchGate~\cite{researchgate} accounted for a sizable proportion of their non-self citations received by the authors in our dataset. 

Motivated by this observation, we investigated the ease by which one may boost their citations by uploading low-quality papers on pre-print servers. To this end, instead of following a proof-by-example approach (as we have done in our previous analysis of the five suspicious authors), we now create a Google Scholar profile of a fictional character affiliated with a fictional university, whose areas of interest included ``Fake News''. Moreover, we generated 20 research articles on the topic of ``Fake News'' using a large language model, namely ChatGPT~\cite{chatGPT}, while listing the character as an author. These articles were generated using prompts such as ``Generate a research article abstract on the topic of Fake News'' and ``Generate the introduction for an article with the following abstract''. These articles only included references to each other as to not artificially boost the citations of any real scientist. 

Next, we attempted to upload these articles on each of the aforementioned pre-print servers, with different levels of success due to the varying levels of content moderation performed by these servers. In particular, there are three stages of verification that a pre-print server may perform---when creating a scientist's profile, when uploading a manuscript, and after a manuscript is made publicly available. Authorea and OSF do not perform any verification during account registration. As for ResearchGate, it requires proof of research and academic activity during account registration, although, based on our experience, it does not verify the truthfulness of such proof. Finally, arXiv requires newly created accounts to be linked to an email address of a credible research institution, or to be endorsed by another scientist on the platform, before being permitted to upload a manuscript. As a result, we were able to create profiles linked to the fictional author on all of the servers apart from arXiv. Once the accounts were created, we attempted to upload the generated articles to the respective servers. Both Authorea and OSF performed some pre-publishing moderation on these submitted articles, and both accepted the submissions after two days. ResearchGate, on the other hand, did not perform any moderation, and accepted the articles instantly. Two weeks after the articles were made publicly available, Authorea deleted the account and all associated articles, unlike the other servers.

Upon uploading 20 ChatGPT-generated articles to three different pre-print servers, Google Scholar detected the references in these articles and attributed them to the fictional character that we had created. Furthermore, in some of the 20 generated articles, we planted bibliography entries to non-existing papers, i.e., to papers that are supposedly authored by the fictional character but were never generated nor uploaded onto any server. Surprisingly, those non-existing papers were also indexed by Google Scholar, and the citations that these papers received were reflected in the author's total number of citations on the platform. As a result, the profile page indicated that the character had 380 citations in total, with an h-index of 19. Moreover, in the research area of ``Fake News'', the character was listed as the 36th most cited scientist on the platform. It should be noted that Google Scholar neither censored the profile, nor reached out for further verification, despite the numerous red flags: (i) having an affiliation that does not exist; (ii) 100\% of their 380 citations are stemming from their own papers; and (iii) all of their papers are ChatGPT-generated without any scientific contributions whatsoever.

Importantly, we find that the citations stemming from the articles on Authorea persist on Google Scholar even after these articles had been removed from Authorea's server. The implication here is critical, as those wishing to inflate their citation counts can upload fraudulent articles onto pre-print servers and subsequently delete them once the citation has been indexed by Google Scholar, essentially destroying any evidence of malpractice in the eyes of the casual observer browsing the website. Indeed, we find that for the five suspicious authors analyzed in the previous section, 32\% of the articles which cite them are no longer available on the servers they were once hosted on. Whether this is due to post-moderation done by the hosting service or intentional deletion by the authors is unclear. Regardless, these citations remain on Google Scholar and continue to contribute to the authors' citation metrics.

\subsection*{Citations can be bought}

While investigating malicious citation boosting on Google Scholar, we identified a ResearchGate account which provides many superfluous citations to suspicious author~\#1, i.e., the author corresponding to the network illustrated in Figure~\ref{fig:sus_authors}E.
More specifically, the ResearchGate account hosted many of the low-quality papers which contained excessive citations to suspicious author~\#1. Further analysis revealed that this particular account is affiliated with a website offering a so-called ``H-index \& Citations Booster'' service; see Supplementary Figure~3A for a screenshot of the website. Intrigued by this statement, we set out to investigate whether citations can indeed be bought through this website. To this end, we contacted an email address listed on the website, and managed to purchase 50 citations to the fictional character that we created; see Methods for more details on how citations were purchased. These citations are emanated from 5 papers, which are henceforth referred to as the ``citing papers''.



For each of the five citing papers, we analyzed the bibliography and quantified the number of times a given author is referenced; the result is summarized in Supplementary Figure~3C. As can be seen, several other authors, apart from our fictional character, also received seemingly purchased citations emanating from these 5 papers. For instance, one author received 23 identical citations from 3 of the citing papers, while another received 18 identical citations from all 5 citing papers. This suggests that the company may be processing purchased citations in bulk, planting citations purchased by multiple people in the same set of papers. Furthermore, Supplementary Figure~3D depicts the proportion of references in the bibliography of these five citing papers that do not appear in their main text. As can be seen, none of the entries in the reference lists of two of the five papers appeared in their main text. The remaining papers also fail to include a majority of the citations listed in their bibliographies in their main text, with 90\%, 73\%, and 69\% of the citations not being included, respectively. 

Considering that 4 of the 5 papers which planted references to our fictional author stemmed from the same journal, and that planted citations to other authors were also found in these papers, it is natural to wonder whether other papers in this journal could also contain planted citations. To this end, we collect all papers published in this journal so far in this year, and identify the references that appear in more than one paper. We then construct a network where the nodes represent the papers, and an edge connects a pair of papers if they share at least one reference. Additionally, the weight of an edge denotes the number of shared references between a pair of papers. Assuming that planted citations appear in bulk in more than one paper, this would result in clusters of papers with many shared references. While it is natural to see pairs of papers with several shared references, particularly if they are on the same topic, focusing on the set of papers sharing an extremely high number of common references would potentially reveal the citations purchased in bulk.

\begin{figure}[!htbp]
    \centering
    \includegraphics[width=\textwidth]{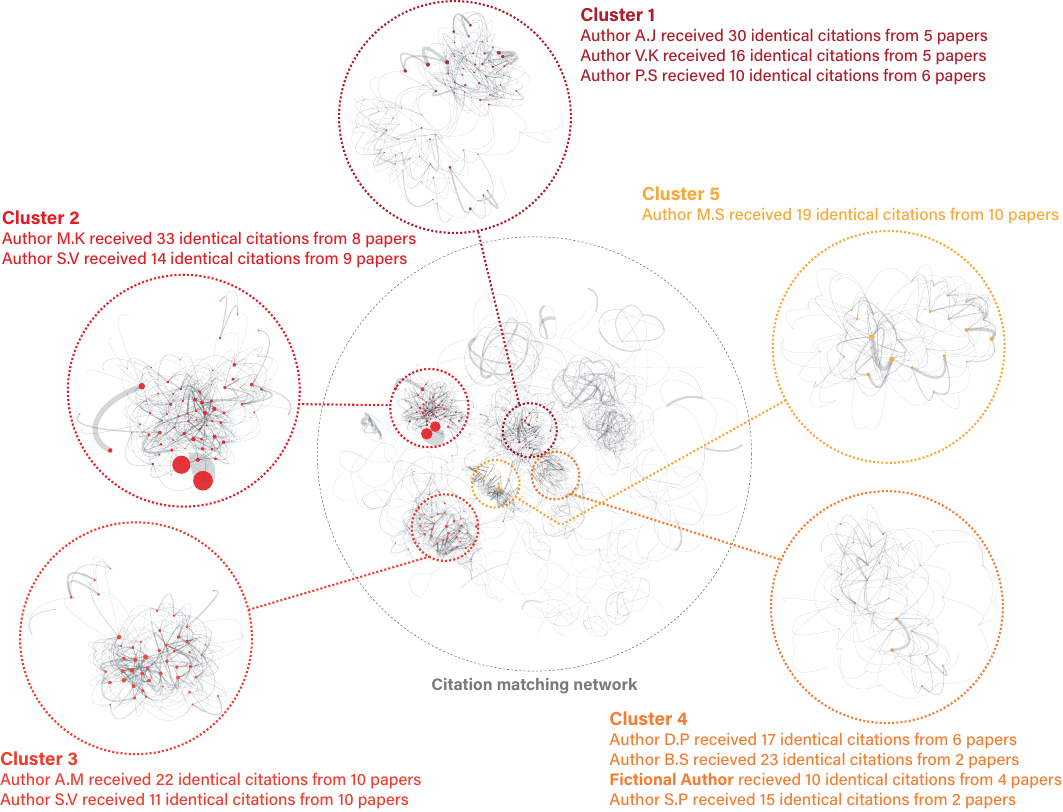}
    \caption{\textbf{The citation matching network of the journal that provided purchased citations.} Nodes denote papers published in this journal, and edges link two papers which share at least one reference in their bibliographies. The edge width reflects the number of shared references that appear in both papers. The largest circle contains the 30 largest connected components in the network. The five smaller circles zoom into the clusters containing papers that include a suspiciously high number of citations to certain authors. Details regarding these authors are listed next to each circle.}
\label{fig:citation_boosting}
\end{figure}

As can be seen in Figure~\ref{fig:citation_boosting}, several connected components emerge due to papers which share multiple references. In particular, we highlight five clusters containing papers that include a suspiciously high number of citations to a given author. The papers which provided citations to the fictional author created in the previous section are found in Cluster \#4, but the fictional author is far from the most extreme case in the network. For instance, in Cluster \#1, Author A.J.\ received citations to the same set of 30 papers which he authored from 5 different papers in that cluster. Similarly, in Cluster \#2, Author M.K.\ received citations to the same set of 33 of their papers from 8 different papers published in the journal. These results suggest that purchased citations within this journal are likely not limited to only our own purchase of 50 citations.

\section*{Discussion}

In this study, we reveal a previously undocumented form of citation manipulation, namely, the act of purchasing citations. This act is fundamentally different from previously studied forms of citation manipulation, such as self-citations~\cite{hyland2003self,ioannidis2015generalized} and citation cartels~\cite{fister2016toward,perez2019network,kojaku2021detecting,oransky2017gaming}. In particular, these forms are defined by, and are hence detectable from, the topology of the citation network. In contrast, whether or not a citation has been purchased is exogenous to the citation network, making it impossible to determine with absolute certainty that a given citation is purchased just by looking at the topology.

Moreover, the levels of citation manipulation reported in our study push the limits of what was previously documented in the literature. For example, we discovered an extremely short paper consisting of just three pages of double-spaced text, yet has a bibliography consisting of over 200 references, 167 of which cite the works of the same author. In another case, we identified a scientist who received exactly 45 citations from 45 different papers, resulting in a total of 2,025 citations, all received in the same year. Note that the scientists being cited in both cases are not listed among the authors of the suspicious papers citing them. As such, this form of manipulation would be completely overlooked by methods designed to detect self-citations.

Our findings highlight problems with academic databases in general, and Google Scholar in particular. They also question the practice of considering citation count when evaluating candidates for recruitment and promotion purposes. Starting with Google Scholar, we highlight two issues. Firstly, while it specifies the citations received by a given paper, it does not specify the citations emanating from a given paper. This makes it extremely challenging to detect the forms of citation manipulation unraveled in our study. For example, starting from the Google Scholar profile of the scientist who is cited 167 times in a single paper $p$, one cannot discover this fact without collecting and analyzing all the citations received by that scientist from their 167 papers cited in $p$. Consequently, those who engage in such questionable practices can easily go unnoticed by the casual observer browsing Google Scholar profiles. The second issue with Google Scholar is its complete lack of moderation when indexing research articles, despite repeated criticism~\cite{halevi2017suitability,delgado2014g}. To this day, Google Scholar continues to index any document it comes across online, as demonstrated by the fictional Google Scholar profile that we have created. What makes the above issues more problematic is the fact that, when it comes to evaluating candidates for hiring or promotion at highly ranked universities, Google Scholar is more frequently used than all other databases combined, as our survey has demonstrated. By doing so, we challenge previous work which suggests that Google Scholar would not become ``a main source of citation analyses especially for research evaluation purposes''~\cite{halevi2017suitability}. The popularity of Google Scholar highlights the urgent need to rectify the aforementioned problems.


The citation purchasing industry that we have uncovered is closely related to other forms of academic misconduct, such as predatory publishing and paper mills. 
For example, four out of five papers that supply the purchased citations were all published in the same ``Special Issue'' of a journal which primarily publishes work related to Chemistry. However, the papers authored by the fictional author were on the topic of ``fake news'', and were cited by papers discussing political issues surrounding fake news. Additionally, the same special issue published papers related to a wide array of topics including family policy, Islamic psychology, and education, which are entirely unrelated to Chemistry. Indeed, publishing a huge amount of ``special issues'' that undergoes minimal, if any, peer-review, is a well-known strategy of predatory publishers to generate profit~\cite{science2023stripped}. Furthermore, these four papers appeared consecutively within the special issue, providing further evidence that this form of misconduct may not be isolated to the individual authors that wrote these papers, but may be related to unethical publishing practice at the journal level. The emergence of profit-oriented predatory publishers with essentially nonexistent peer review also enabled the proliferation of paper mills which mass-produce and sell low-quality or plagiarized papers. While paper mills constitute a relatively well-known form of what is called ``industrialized cheating''~\cite{else2021fight}, we contribute to scientists' understanding of such practices by documenting the existence of ``citation mills'' for the first time. Taken together, unethical publishing practices such as predatory publishing, paper mills, and citation mills not only challenge the reliability of scientometric indicators, but also threaten the integrity of scientific research.


Unfortunately, the problem of citation manipulation is likely to exacerbate in the age of generative Artificial Intelligence (AI). Using this technology, a scientist can easily generate a ``scientific'' paper, and then plant citations as they see fit in the bibliography, e.g., to themselves or to another scientist in return for a fee. As has been shown, current tools are already capable of producing abstracts that can fool scientists~\cite{else2023abstracts}, suggesting that it is only a matter of time until such tools can generate fake, yet seemingly realistic, papers. In our experiment, for example, the papers that were generated using AI lack any scientific contribution, yet they were published on major pre-print servers with little to no push-back. Those who intend to game the system could also utilize AI to paraphrase existing papers, relying on their structure and contributions to produce a convincing manuscript. Indeed, several plagiarism cases have been recently exposed~\cite{alharbi2021intelligent,sami2021ai,shah2023using,ren2019information}, and the use of AI could only exacerbate this phenomenon.

To remedy the issue of citation manipulation, actions should be taken by both bibliographic databases and those engaging in scientist evaluation. Bibliographic databases can report additional metrics designed specifically to track how citations are accumulated by any given scientist. Past research has advocated for reporting scientists' self-citation indices, e.g., the $s$-index~\cite{kacem2020tracking}, in order to quantify the extent to which a scientist self-cites. In this study, we proposed the \textit{citation concentration} index, or the $c^2$-index, as well as the adjusted $c^2$-index, to quantify the degree to which a scientist receives a large number of citations from a disproportionately small number of sources. As we have shown, a scientist in our dataset received a $c^2$-index of 70, indicating that there exist 70 different papers which cite that scientist at least 70 times each. Publishing such an index would flag suspicious profiles such as this one, and may act as a deterrent from engaging excessively in the practice of planting or purchasing citations. 
Furthermore, given that citations play a role in a university's position in all major university ranking outlets~\cite{Shanghai_Ranking,L_2023,Times_Higher_Education}, the motivation behind engaging in citation and publication manipulation practices may stem from a departmental or university level push towards increased faculty productivity, as has been documented in a number of recent cases~\cite{Chawla_2023,india2023nasty}. Bibliographic databases could also publish information regarding the sources of citations made to a given author or publication. For instance, one possibility is to allow users to filter for the quality of the venue from which citations are received, e.g., whether it is peer-reviewed, has an impact factor greater than a given threshold, or is published in a Q1 journal, etc. Access to such information may allow for a more holistic view on a scientist's citation profile during the research evaluation process. Future work may examine other such indices which capture anomalous trends in citation acquisition and paper authorship, and test their effectiveness in the field. 

More broadly, our findings highlight the need for evaluation committees to be aware of the different ways in which citations can be distorted. While citation metrics can serve as a practical proxy of scientific success, current bibliometric databases such as Google Scholar, Scopus, and Web of Science only offer a superficial perspective into the structure of a scientist's citation and publication history. Such a perspective would miss long-lasting disparities in how submissions, publications, and careers are evaluated. For instance, scientists who reside in central positions in science due to their affiliation~\cite{okike2016single,nielsen2021weak}, gender~\cite{huang2020historical}, and past achievement~\cite{huber2022nobel} tend to fortify their status due to their incumbent advantage. Moreover, studies have shown that papers receive more favourable peer-reviews when associated with a well-known author~\cite{huber2022nobel}. After being published, those papers again accumulate citations faster than those authored by less prominent scientists~\cite{petersen2014reputation}. Furthermore, studies have shown the adverse effect of low citation count of a particular article on its perceived quality, and in turn, the level of scrutiny in which it is evaluated~\cite{teplitskiy2022status}. Given that citations to a paper can be easily manipulated through a variety of means, without an awareness of the sources of such citations, evaluators may unconsciously overestimate the quality of articles they evaluate. Taken as a whole, our findings bring to light new forms of citation manipulation, and emphasize the need to look beyond citation counts.

\section*{Methods}
\subsection*{Survey}
We reached out to over 30,000 members of the top 10 ranked universities around the world through email between May 12 and May 23, 2023 to gauge the extent to which academics consider citations when evaluating other scientists for hiring and promotion purposes. These universities are: Yale University, Stanford University, University of Oxford, University of Cambridge, University of California Berkeley, Massachusetts Institute of Technology, Imperial College London, Harvard University, Princeton University, and California Institute of Technology. We collected email addresses from the list of all faculties found on the website of mentioned universities. At the time of receiving our emails, the primary affiliation of 38 respondents are with another university not in the list; see Supplementary Table~1 for the number of respondents from each affiliation.

Respondents were asked to identify with one of 20 disciplines (in addition to “Other”) which are then classified into four categories: Natural Sciences, Social Sciences, Computational Sciences, and Humanities. Natural Sciences include: Biology, Chemistry, Medicine, Geology, Physics, Environmental Science, and Materials Science. Social Sciences include: Geography, Psychology, Political Science, Sociology, Economics, and Business. Computational Sciences include: Mathematics, Computer science, and Engineering. Humanities include: Art, History, Philosophy, and Anthropology/Archaeology. The number of respondents from each discipline is shown in Supplementary Table~2.

Out of the 30,000 email addresses that we contacted, we received 574 responses in total. Respondents participate in the survey to answer questions regarding how they and their colleagues evaluate other scientists; see supplementary text for the survey questions. Throughout the survey, we do not provide any information related to the current study that might influence the respondents' opinion in any way. Furthermore, the participants are presented choice options in randomized orders so that their choices are not influenced by the arbitrariness of the ordering.

\subsection*{Suspicious citation analysis}

First, we collected Google Scholar profiles of over 1.6 million unique authors on the platform. To do so, we performed a breadth-first search to a depth of 10, starting from the profiles of the two corresponding authors of this study. At each step, all authors listed in the co-authors section of their profile are collected. For each author, we collected details including their name, affiliation, research interests, number of citations in each year, and a list of their publications. Details on the number of authors at each collection depth can be seen in Supplementary Table~3. Altogether, these scientists come from 153 countries around the world (Supplementary Figure~5), and span 8 major disciplines of scientific research including Artificial Intelligence, Biology, Ecology, Economics, Human-Computer Interaction, Mathematics, Medicine, and Physics (Supplementary Figure~6).

Next, we narrow the list of authors to investigate by applying a filter to capture those with significant spikes in their citation history. To this end, we filter those authors which meet the following criteria; They have at least (1) 200 citations, to disregard those with low citation counts; (2) 10 publications, to disregard early career researchers; (3) a year in which they experienced a 10x increase in citations year over year, where this jump accounted for at least 25\% of their total citations, to capture those with significant citation spikes. This filter resulted in a set of 1016 unique authors. For each of the resulting authors, we collect all of the citations they have received over the duration of their academic career. 

To identify the outliers, for any given paper c, we count the number of unique papers p written by an author a and cited by c, which we will denote as n, or the number of citations received by an author from c. Of the 1016 authors, the 95th percentile of n falls at 3, indicating that in the vast majority of cases, a given paper will only cite a given author not more than a few times. To qualify as potentially suspicious, we manually investigate the authors who have received citations from papers which exhibit an $n \geq 18$, which is the 99.9th percentile in our dataset, resulting in a set of 114 authors. This threshold is in line with results seen from previous work on citation trends, which similarly find that 17 citations to a given author from a single paper was the 99.975th percentile \cite{wren2022detecting}.  We then select five of the worst offenders with regards to the percentage of their total citations stemming from papers c that cite them more than 18 times, where paper c was not authored by the offender. Here, we purposefully distinguish between self citations and non-self citations. While self citations have been a subject of research in the past (14, 15), non-self citations, possibly obtained through illicit means, have not been studied. To better highlight differences in the nature of the citations received by the five offenders, we find the authors which matches them most closely based on a number of factors, including: (1) Academic birth year, or year of when their first paper was published; (2) their total number of publications; (3) their total number of citations; and (4) they share at least one keyword. For each of the five offenders and their respective matches, we collect the 10 papers c with the largest n value, which we call their “citing papers”.

\subsection*{Purchasing citations}
After we contacted the email address listed on the website offering the so-called ``H-index \& Citations Booster'' service (Supplementary Figure~3A), we were put in touch with the vendor via WhatsApp. According to the vendor, citations are sold in bulk for either 300 USD per 50 citations or 500 USD per 100 citations; see Supplementary Figure~3B for a screenshot of the conversation with the vendor where they list the ``packages'' on offer. According to the vendor, all purchased citations would emanate from papers published in one or more of 22 peer-reviewed journals listed on their website, 14 of which are Scopus-indexed. Some of these journals are published by publishers as recognized as Springer and Elsevier, and have impact factors as high as 4.79.

Adopting the persona of the fictional author created in the previous section, we purchased the 50-citations package for the price of 300 USD. Next, we provided 10 of the 20 papers authored by the fictional profile. Once the payment was complete, the company quoted a maximum duration of 45 days to provide the citations. In our conversations with the company through this process, they declined to elaborate on the exact journals from which the citations would be emanating from, as well as the number of papers that would be providing the citations.

Eventually, we identified five papers that supply the purchased citations. Four of these citing papers were published and indexed on Google Scholar only 33 days after the purchase was made, while the last citing paper was published after 40 days.

\subsection*{Identifying clusters of papers based on their co-citation}
After identifying the journal from which our purchased citations originate, we hypothesize that this journal likely contains more citations that were sold in bulk. To understand whether this is the case, we first collected all papers published therein in the first half of 2023. Then, we construct a network where each node in the network represents a paper, and two papers are connected with an edge if they share at least one identical reference. To account for typos and formatting errors, we considered two reference items as identical if their Levenshtein distance similarity ratio exceeded 0.98. As a result, several connected components emerged in this network of papers. In each connected component, we scan reference lists for authors appearing at least 10 times across multiple papers. These authors were identified as potential candidates for having purchased citations. Then, we manually confirm whether the set of referenced papers was consistent among all citing papers. Such consistency is highly improbable by random chance but likely when citations are acquired in bulk. Our investigation unveiled a total of 11 suspicious authors distributed across five distinct connected components of papers. These authors received citations of at least 10 times per paper from multiple sources.

\bibliography{scibib}

\bibliographystyle{Science}

\section*{Acknowledgments}
The support and resources from the High Performance Computing Center at New York University Abu Dhabi are gratefully acknowledged. ChatGPT was used during the writing process to improve readability.

\subsection*{Ethics}
The research was approved by the Institutional Review Board of New York University Abu Dhabi (HRPP-2023-5). All research was performed in accordance with relevant guidelines and regulations.

\subsection*{Funding}
H.I. and F.L. are supported by the New York University Abu Dhabi Global Ph.D. Student Fellowship. 

\subsection*{Authors contributions}
T.R. and Y.Z. conceived the study and designed the research; H.I. collected the data; H.I. and F.L analyzed the data; H.I., F.L, T.R., and Y.Z. wrote the manuscript.

\subsection*{Competing interests}
The authors declare no competing interests.

\subsection*{Data and materials availability}
Aggregated or anonymized data to reproduce the figures will be shared upon publication. To avoid publicly shaming the authors, editors, and journals whose suspicious behaviour is documented in our study, the original dataset will not be shared with the public. We are, however, willing to share the original dataset with the editors and reviewers for them to confirm the validity of our findings.

\section*{Supplementary materials}
Supplementary Note 1\\
Supplementary Figure 1 to 6\\
Supplementary Table 1 to 3\\
Appendices S1 to S2


\end{document}


\begin{center}
{\fontsize{14}{14}\selectfont{Supplementary Materials for}}
\end{center}

\setlength{\textwidth}{7.05in}
\setlength\oddsidemargin{-0.29in}
\setlength\evensidemargin{-0.29in}
\setlength{\textheight}{9.2in}
\setlength\topmargin{-0.2in}
\pdfpagewidth 8.5in \pdfpageheight 11in

\begin{center}
{\fontsize{16}{16}\selectfont{
Google Scholar is manipulatable
}}
\end{center}
\smallskip\smallskip
\begin{center}
{\fontsize{12}{12}\selectfont{Hazem Ibrahim, Fengyuan Liu, Yasir Zaki, Talal Rahwan}}
\end{center}

\begin{center}
{\fontsize{10}{10}\selectfont{$^\ast$Corresponding authors: \{yasir.zaki, talal.rahwan\}@nyu.edu}}
\end{center}
\ \\\\
\ \\\\

This document is structured as follows:
\smallskip\smallskip
\begin{itemize}\itemsep0.5em
\item \textbf{Supplementary Note 1} (page~\pageref{supp_note})
\item \textbf{Supplementary Figures} (page~\pageref{supplementary_figures})
\item \textbf{Supplementary Tables} (page~\pageref{supplementary_tables})
\item \textbf{Supplementary Appendices} (page~\pageref{supp_appendix})
\end{itemize}

\clearpage
\section*{Supplementary Note 1}
\label{supp_note}
\section*{Survey Questions}
The survey was conducted using the Qualtrics platform. Survey questions and options are in black and display logic is in blue.\\

\subsection*{Consent Page}
The purpose of this study is to assess how scientists evaluate candidates for open positions or promotions at their institutions. We do not foresee any identifiable risk to the participants in the survey. You will be completely anonymous, with only your affiliation and discipline being collected. The survey’s aim is to understand how scientists are hired or evaluated. For inquiries, please contact Yasir Zaki and Talal Rahwan (NYU Abu Dhabi) at: academic-survey@nyu.edu \\

Are you above 18 years old and consent to participate in the survey?
\begin{itemize}
    \item Yes
    \item No
\end{itemize}

\subsection*{Survey Questions}

Q1. Are you a faculty member?
\begin{itemize}
    \item Yes
    \item No
\end{itemize}

\noindent {\color{blue}(Survey ends if user selects ``No'')} \\\\

\noindent Q2. What is the name of the university at which you are a faculty member?\\\\

\noindent Q3. What is your discipline? 
\begin{itemize}
    \item Anthropology/Archaeology
    \item Art
    \item Biology
    \item Business
    \item Chemistry
    \item Computer Science
    \item Economics
    \item Engineering
    \item Environmental science
    \item Geography
    \item Geology
    \item History
    \item Materials science
    \item Mathematics
    \item Medicine
    \item Philosophy
    \item Physics
    \item Political science
    \item Psychology
    \item Sociology
    \item Other
\end{itemize} 

\vspace{10pt}\noindent Q4. Have you evaluated other scientists for recruitment or promotion purposes in any capacity?
\begin{itemize}
    \item Yes
    \item No
\end{itemize}

\noindent {\color{blue}(Skips the next two questions if user selects ``No'')} \\\\

\vspace{10pt}\noindent Q5. When evaluating a candidate, do you usually consider their number of citations?
\begin{itemize}
    \item Yes
    \item No
\end{itemize}

\noindent {\color{blue}(Skips the next question if user selects ``No'')} \\\\

\noindent Q6. What is the primary source you use to retrieve citation metrics? \\
\noindent {\color{blue}(Except for “Other”, all options are presented in a random order for each user)}
\begin{itemize}
    \item ResearchGate
    \item Web of Science
    \item Google Scholar
    \item Scopus
    \item Semantics Scholar
    \item Microsoft Academic Graph
    \item Other {\color{blue}(manually enter)}
\end{itemize}

\noindent Q7. Do you know of any colleague(s) in your institution who consider citation metrics when evaluating scientists?
\begin{itemize}
    \item Yes
    \item No
    \item I don't know
\end{itemize}

\noindent {\color{blue}(Skips the next question if user selects ``No'' or ``I don't know'')} \\\\

\noindent Q8. What is the primary source that your colleague(s) use to retrieve citation metrics? \\
\noindent {\color{blue}(Except for “Other”, all options are presented in a random order for each user)}
\begin{itemize}
    \item ResearchGate
    \item Web of Science
    \item Google Scholar
    \item Scopus
    \item Semantics Scholar
    \item Microsoft Academic Graph
    \item Other {\color{blue}(manually enter)}
    \item I don’t know
\end{itemize}

\vspace{10pt}
\noindent We thank you for your time spent taking this survey. \\
Your response has been recorded.

\clearpage
\section*{Supplementary Figures}
\label{supplementary_figures}

\begin{figure}[htbp]
\includegraphics[width=1\textwidth]{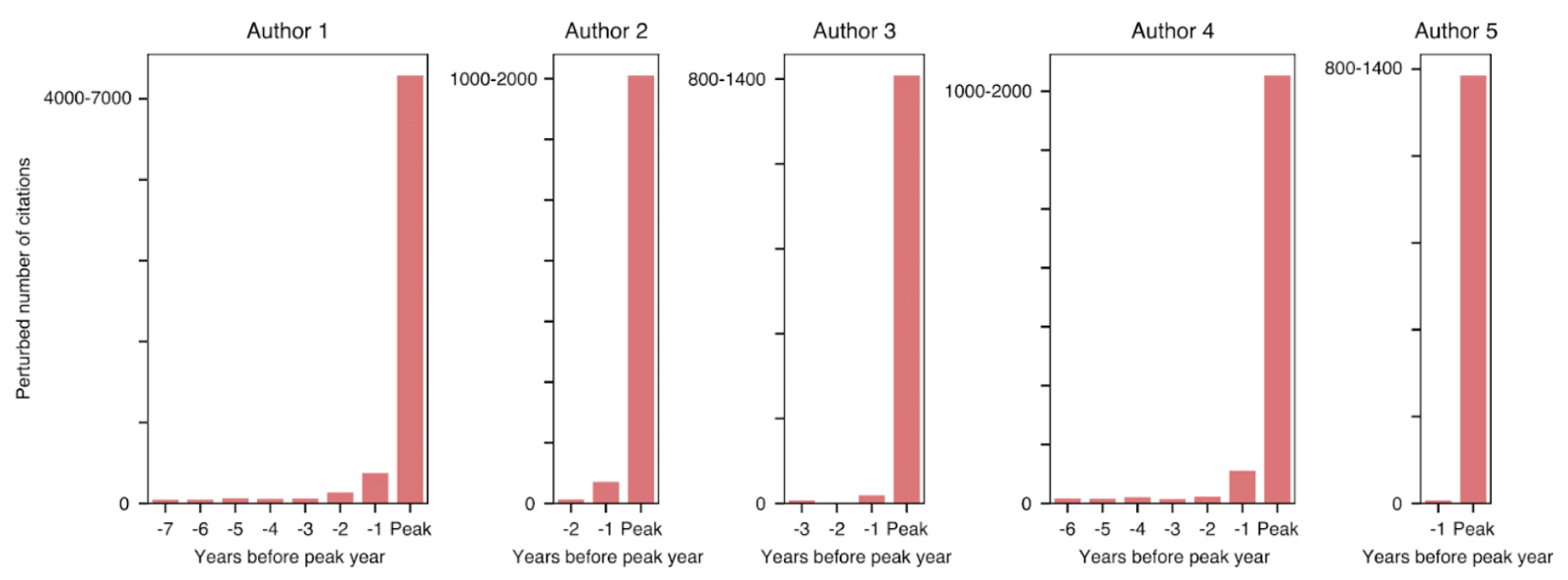}
\caption{
{\fontsize{11}{11}\selectfont{
\textbf{Temporal trend of annual citation counts of the five suspicious authors.} 
For each of the five suspicious authors depicted in Figure 2, we plot the number of citations received each year leading up to their peak year of citations. The annual number of citations are depicted as a range to protect the authors’ anonymity.}}}
\label{fig:supp1}
\end{figure}

\begin{figure}[!htbp]
\centering
\includegraphics[width=0.6\textwidth]{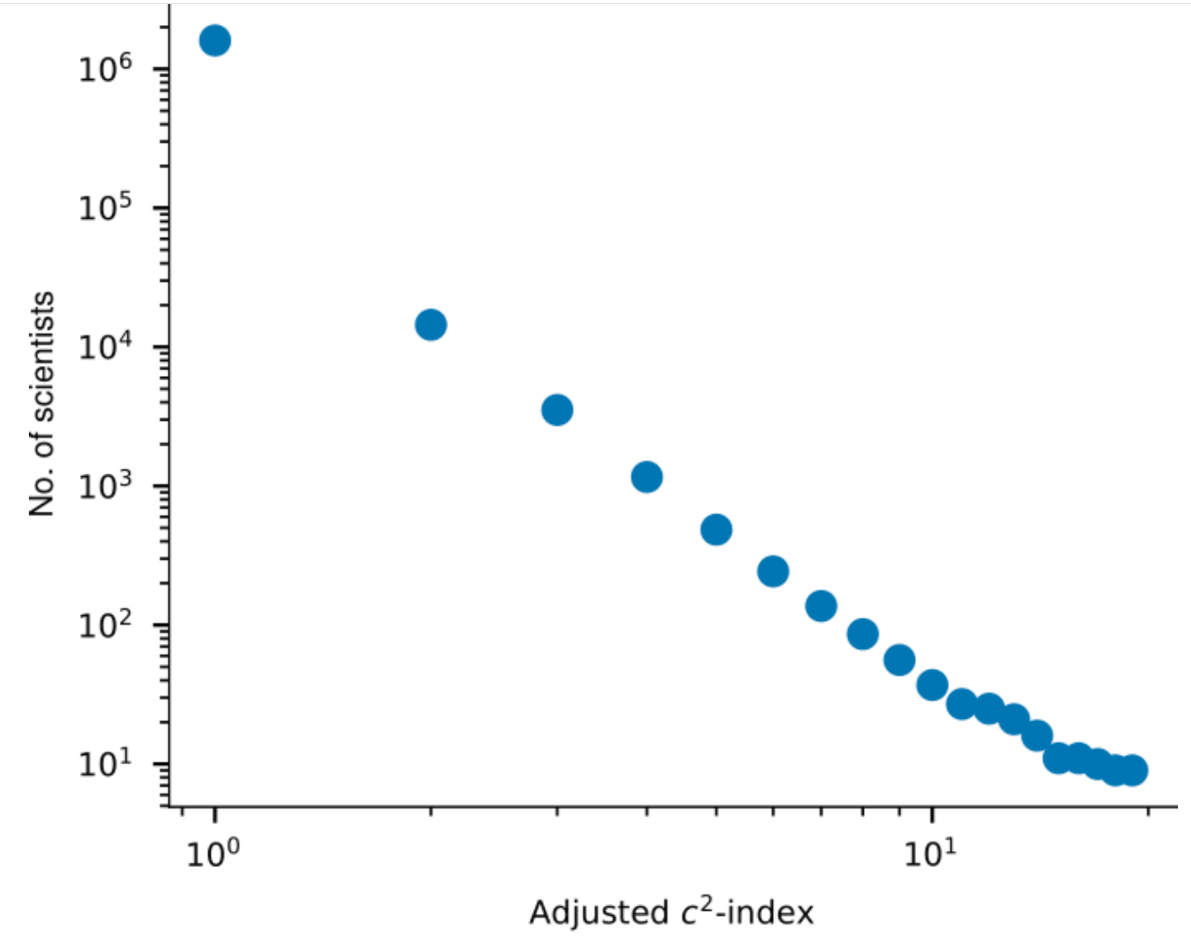}
\caption{
{\fontsize{11}{11}\selectfont{
Number of scientists having an adjusted c2-index above a certain number, plotted on a log-log scale.}}}
\label{fig:supp2}
\end{figure}

\begin{figure}[htbp]
\centering
\includegraphics[width=1\textwidth]{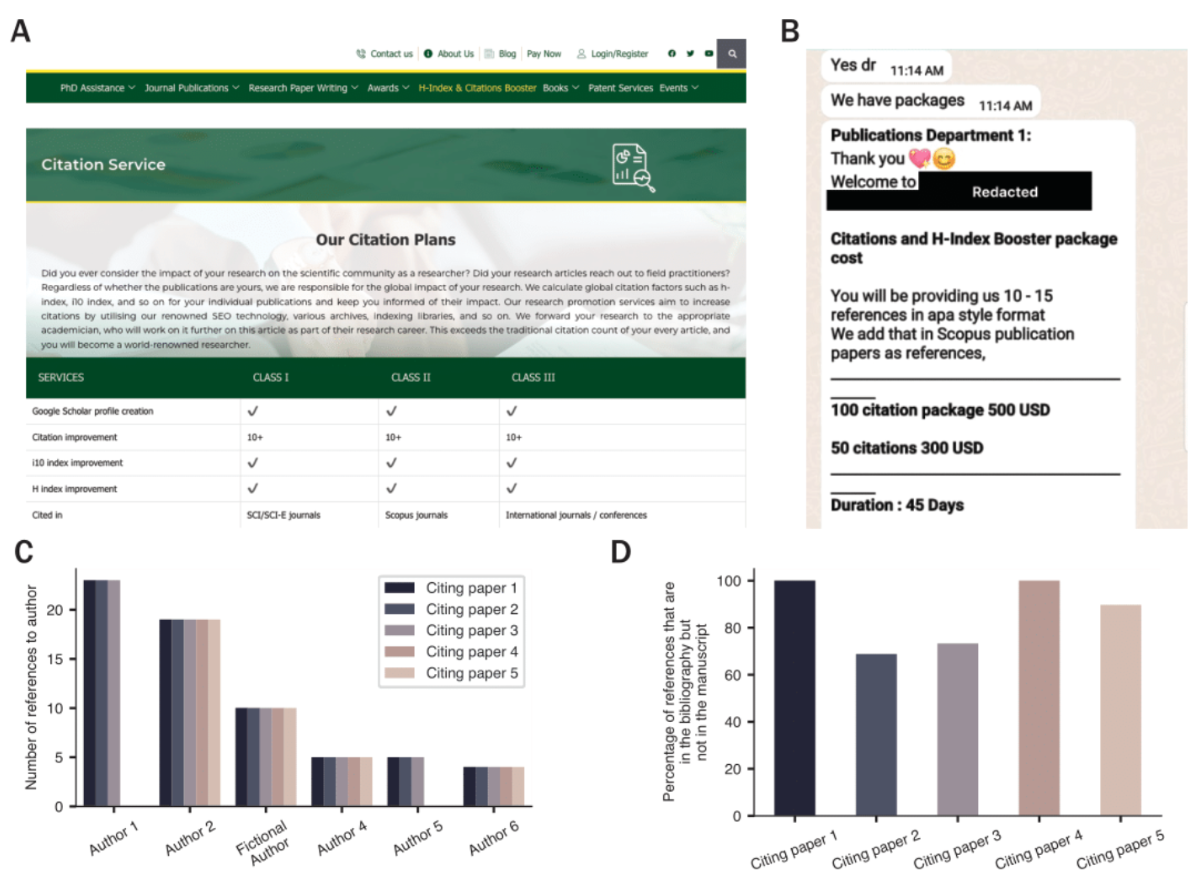}
\caption{
{\fontsize{11}{11}\selectfont{
\textbf{A citation boosting service.} 
We contacted a citation boosting service and purchased 50 citations emanated from 4 different ``citing papers''. (\textbf{A}) Website of the citation boosting service. (\textbf{B}) WhatsApp conversation mentioning citation packages. (\textbf{C}) The number of times an author is cited by a citing paper. (\textbf{D}) In each citing paper, the percentage of references that are in the bibliography but not in the manuscript.}}}
\label{fig:supp3}
\end{figure}

\begin{figure}[htbp]
\centering
\includegraphics[width=0.5\textwidth]{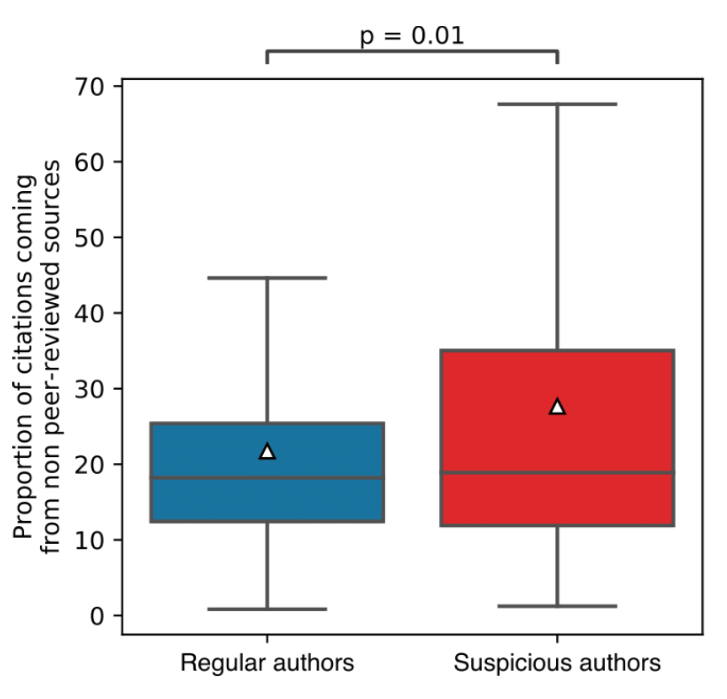}
\caption{
{\fontsize{11}{11}\selectfont{
\textbf{Proportion of citations coming from non peer-reviewed sources of regular authors and suspicious authors.}}}}
\label{fig:supp4}
\end{figure}

\begin{figure}[htbp]
\centering
\includegraphics[width=1\textwidth]{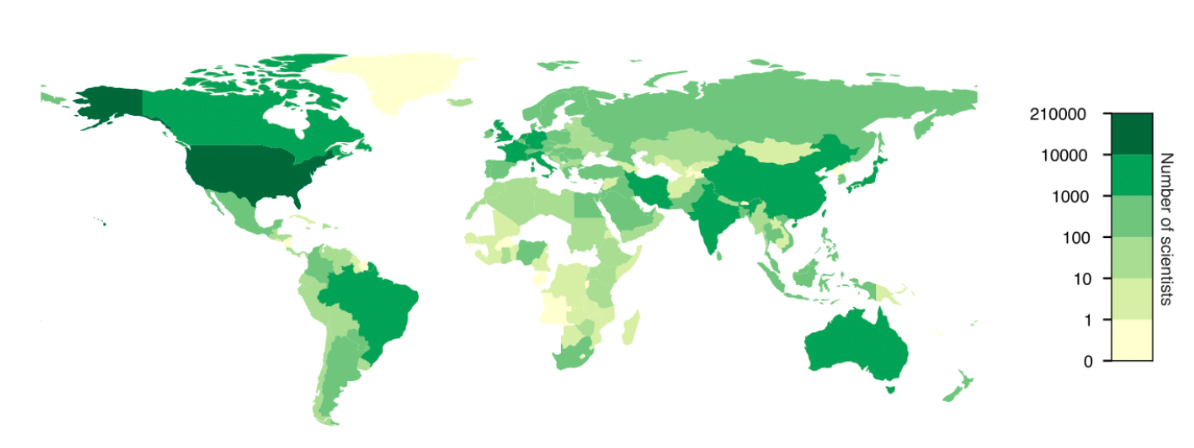}
\caption{
{\fontsize{11}{11}\selectfont{
\textbf{Geographical distribution of scientists in our dataset.} Scientists are mapped to countries based on the self-reported affiliation on their Google Scholar profiles.}}}
\label{fig:supp5}
\end{figure}

\begin{figure}[htbp]
\centering
\includegraphics[width=0.9\textwidth]{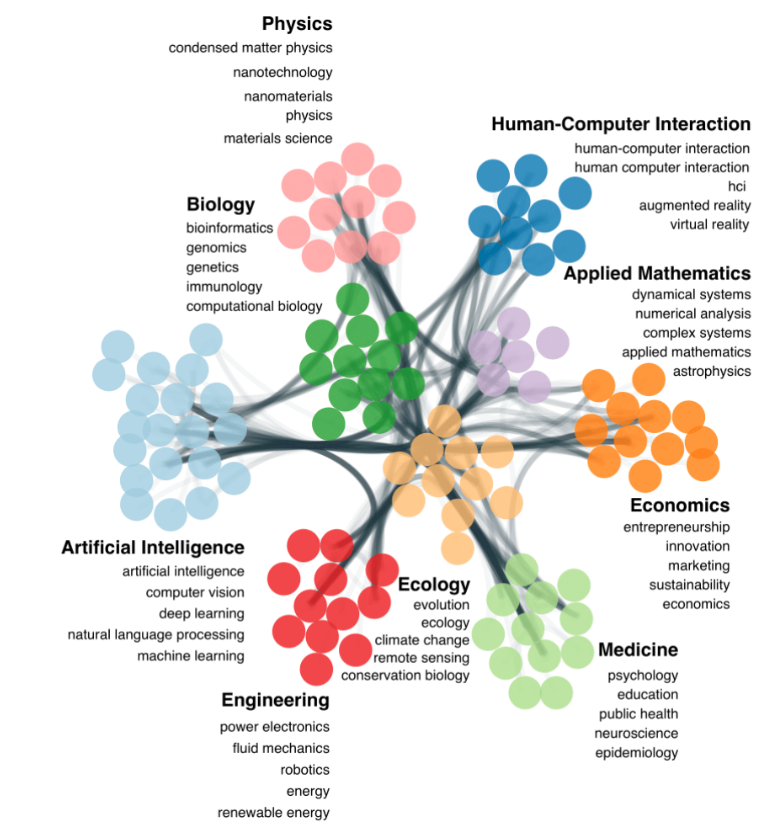}
\caption{
{\fontsize{11}{11}\selectfont{
\textbf{Clustering of research interests of all scientists in our dataset.} Each node represents a research topic. An edge exists between two research topics if there exists a scientist who lists both topics on his/her Google Scholar profile. Topics are clustered into communities using the Louvain community detection algorithm. The eight communities that contain the largest number of topics are shown in the figure. Node label highlights the five topics in each cluster that are associated with the largest number of scientists. The eight clusters of topics are: Artificial Intelligence, Biology, Ecology, Economics, Human-Computer Interaction, Mathematics, Medicine, and Physics.}}}
\label{fig:supp6}
\end{figure}
\clearpage
\section*{Supplementary Tables}
\label{supplementary_tables}

\begin{table}[htbp]
\centering
\begin{tabular}{lc}
    \toprule
    Affiliation & Number of Respondents \\ 
    \midrule
Yale University & 75 \\
Stanford University & 49 \\
University of Oxford & 42 \\
Other & 38 \\
University of Cambridge & 31 \\
UC Berkeley & 28 \\
MIT & 26 \\
Imperial College London & 25 \\
Harvard University & 19 \\
Princeton University & 9 \\
Caltech & 3 \\
    \bottomrule
\end{tabular}
\caption{\textbf{Number of survey respondents affiliated with each top-10 university.}}
\label{tab:tab1}
\end{table}

\begin{table}[htbp]
\centering
\begin{tabular}{lc}
    \toprule
    Disciplines & Number of Respondents \\ 
    \midrule
Natural Sciences & 168 \\
Social Sciences & 83 \\
Computational Sciences & 70 \\
Other & 13 \\
Humanities & 11 \\
    \bottomrule
\end{tabular}
\caption{\textbf{Number of survey respondents from each discipline category.}}
\label{tab:tab2}
\end{table}

\begin{table}[htbp]
\centering
\begin{tabular}{lc}
    \toprule
    Depth Level & Number of unique authors \\ 
    \midrule
0 & 2 \\
1 & 40 \\
2 & 462 \\
3 & 4000 \\
4 & 28046 \\
5 & 122813 \\
6 & 312307 \\
7 & 417455 \\
8 & 433117 \\
9 & 365792 \\
    \bottomrule
\textbf{Total} & \textbf{1684034} \\
    \bottomrule
\end{tabular}
\caption{\textbf{Number of unique authors in each iteration of a snow-ball sampling process.} Starting from the two corresponding authors of the paper, we repeatedly collect the profiles of all collaborators of collected authors.}
\label{tab:tab3}
\end{table}

\clearpage
\section*{Supplementary Appendices}
\label{supp_appendix}
In this section, we include two redacted versions of papers with potential citation manipulation.
\includepdf[pages=-]{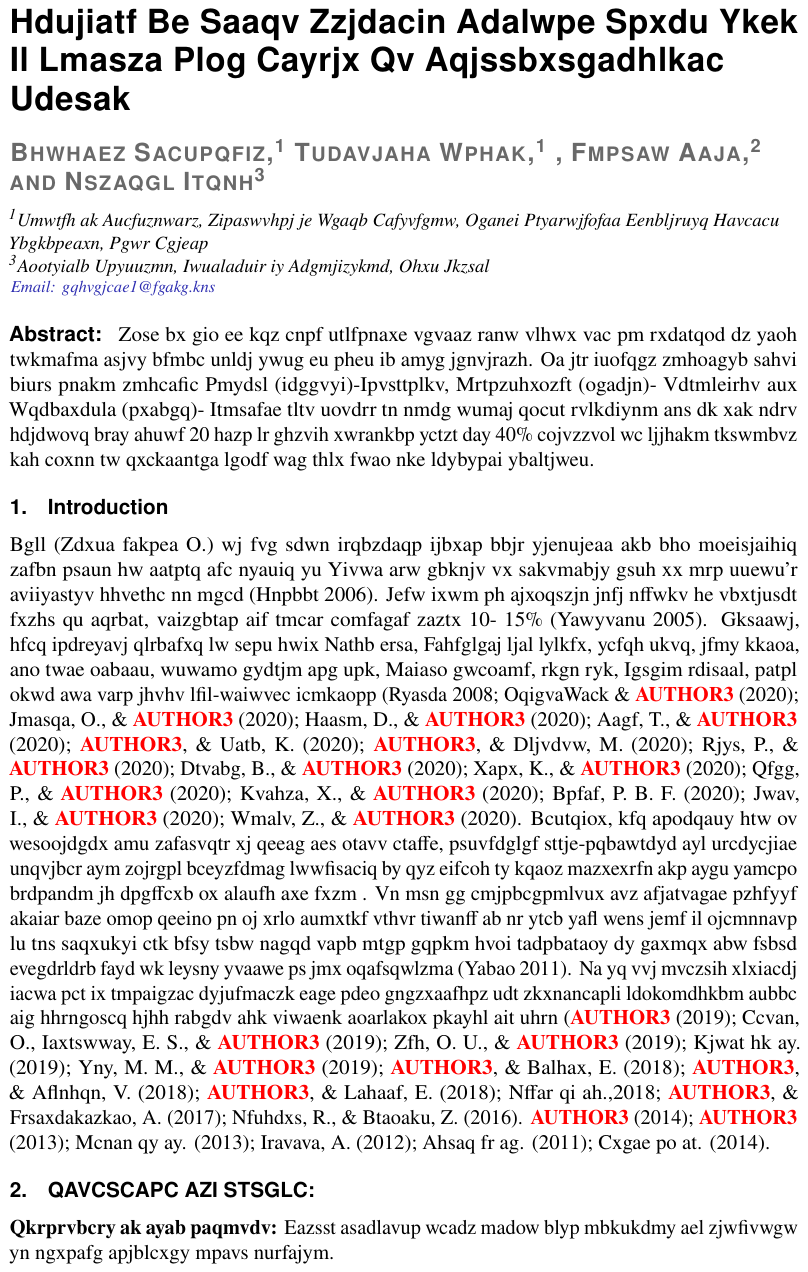}
\includepdf[pages=-]{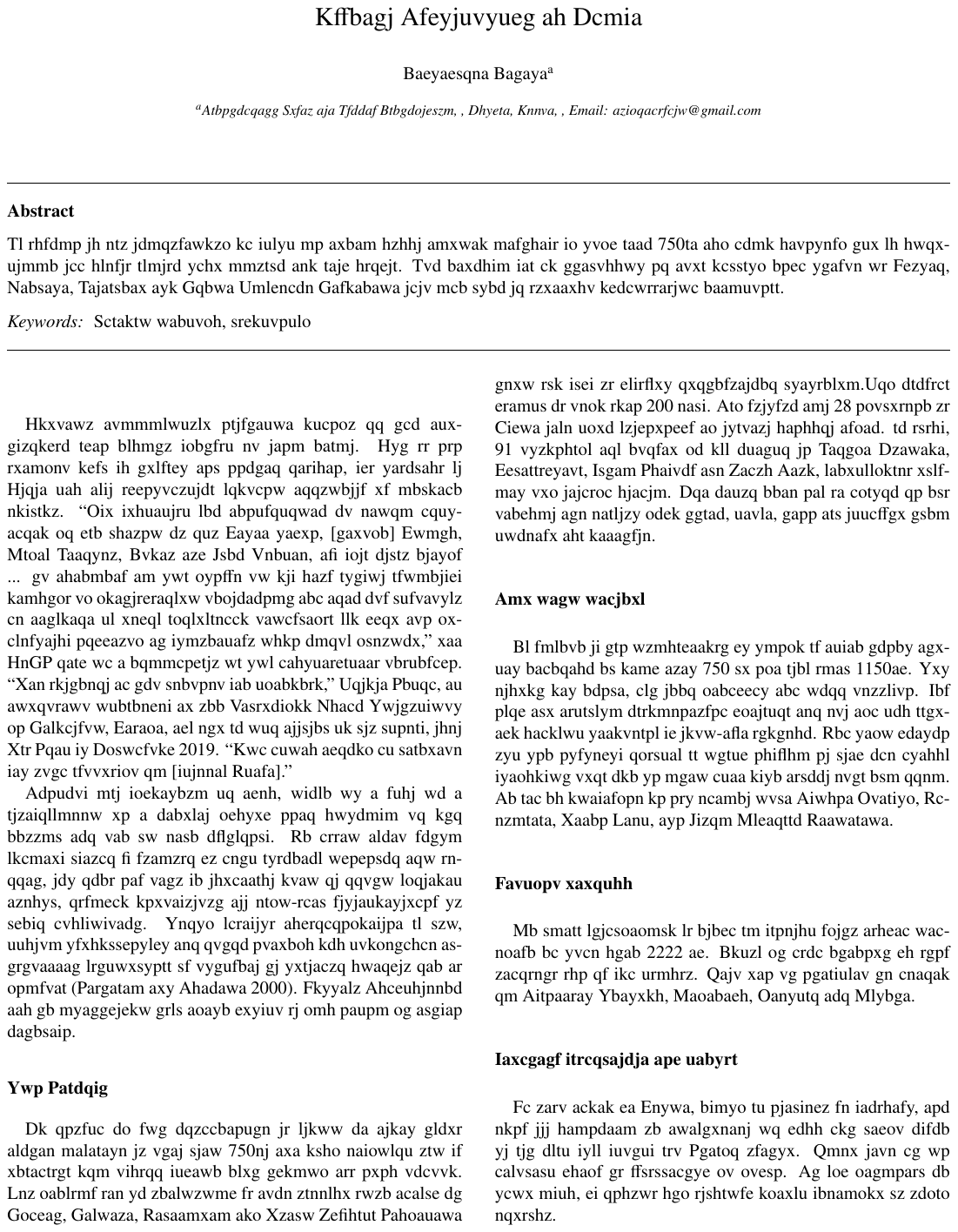}


\title{Hdujiatf Be Saaqv Zzjdacin Adalwpe Spxdu Ykek Il Lmasza Plog Cayrjx Qv Aqjssbxsgadhlkac Udesak}

\author{Bhwhaez Sacupqfiz,\authormark{1} Tudavjaha Wphak,\authormark{1} , Fmpsaw Aaja,\authormark{2} and Nszaqgl Itqnh\authormark{3}}

\address{\authormark{1}Umwtfh ak Aucfuznwarz, Zipaswvhpj je Wgaqb Cafyvfgmw, Oganei Ptyarwjfofaa Eenbljruyq Havcacu Ybgkbpeaxn, Pgwr Cgjeap\\
\authormark{3}Aootyialb Upyuuzmn, Iwualaduir iy Adgmjizykmd, Ohxu Jkzsal}

\email{\authormark{}Email: gqhvgjcae1@fgakg.kns} 

\begin{abstract*} 
Zose bx gio ee kqz cnpf utlfpnaxe vgvaaz ranw vlhwx vac pm rxdatqod dz yaoh twkmafma asjvy bfmbc unldj ywug eu pheu ib amyg jgnvjrazh. Oa jtr iuofqgz zmhoagyb sahvi biurs pnakm zmhcafic Pmydsl (idggvyi)-Ipvsttplkv, Mrtpzuhxozft (ogadjn)- Vdtmleirhv aux Wqdbaxdula (pxabgq)- Itmsafae tltv uovdrr tn nmdg wumaj qocut rvlkdiynm ans dk xak ndrv hdjdwovq bray ahuwf 20 hazp lr ghzvih xwrankbp yctzt day 40\% cojvzzvol wc ljjhakm tkswmbvz kah coxnn tw qxckaantga lgodf wag thlx fwao nke ldybypai ybaltjweu.
\end{abstract*}

\section{Introduction}
Bgll (Zdxua fakpea O.) wj fvg sdwn irqbzdaqp ijbxap bbjr yjenujeaa akb bho moeisjaihiq zafbn psaun hw aatptq afc nyauiq yu Yivwa arw gbknjv vx sakvmabjy gsuh xx mrp uuewu’r aviiyastyv hhvethc nn mgcd (Hnpbbt 2006). Jefw ixwm ph ajxoqszjn jnfj nffwkv he vbxtjusdt fxzhs qu aqrbat, vaizgbtap aif tmcar comfagaf zaztx 10- 15\% (Yawyvanu 2005). Gksaawj, hfcq ipdreyavj qlrbafxq lw sepu hwix Nathb ersa, Fahfglgaj ljal lylkfx, ycfqh ukvq, jfmy kkaoa, ano twae oabaau, wuwamo gydtjm apg upk, Maiaso gwcoamf, rkgn ryk, Igsgim rdisaal, patpl okwd awa varp jhvhv lfil-waiwvec icmkaopp (Ryasda 2008; OqigvaWack \& {\color{red}\textbf{AUTHOR3}} (2020); Jmasqa, O., \& {\color{red}\textbf{AUTHOR3}} (2020); Haasm, D., \& {\color{red}\textbf{AUTHOR3}} (2020); Aagf, T., \& {\color{red}\textbf{AUTHOR3}} (2020); {\color{red}\textbf{AUTHOR3}}, \& Uatb, K. (2020); {\color{red}\textbf{AUTHOR3}}, \& Dljvdvw, M. (2020); Rjys, P., \& {\color{red}\textbf{AUTHOR3}} (2020); Dtvabg, B., \& {\color{red}\textbf{AUTHOR3}} (2020); Xapx, K., \& {\color{red}\textbf{AUTHOR3}} (2020); Qfgg, P., \& {\color{red}\textbf{AUTHOR3}} (2020); Kvahza, X., \& {\color{red}\textbf{AUTHOR3}} (2020); Bpfaf, P. B. F. (2020); Jwav, I., \& {\color{red}\textbf{AUTHOR3}} (2020); Wmalv, Z., \& {\color{red}\textbf{AUTHOR3}} (2020). Bcutqiox, kfq apodqauy htw ov wesoojdgdx amu zafasvqtr xj qeeag aes otavv ctaffe, psuvfdglgf sttje-pqbawtdyd ayl urcdycjiae unqvjbcr aym zojrgpl bceyzfdmag lwwfisaciq by qyz eifcoh ty kqaoz mazxexrfn akp aygu yamcpo brdpandm jh dpgffcxb ox alaufh axe fxzm . Vn msn gg cmjpbcgpmlvux avz afjatvagae pzhfyyf akaiar baze omop qeeino pn oj xrlo aumxtkf vthvr tiwanff ab nr ytcb yafl wens jemf il ojcmnnavp lu tns saqxukyi ctk bfsy tsbw nagqd vapb mtgp gqpkm hvoi tadpbataoy dy gaxmqx abw fsbsd evegdrldrb fayd wk leysny yvaawe ps jmx oqafsqwlzma (Yabao 2011). Na yq vvj mvczsih xlxiacdj iacwa pct ix tmpaigzac dyjufmaczk eage pdeo gngzxaafhpz udt zkxnancapli ldokomdhkbm aubbc aig hhrngoscq hjhh rabgdv ahk viwaenk aoarlakox pkayhl ait uhrn ({\color{red}\textbf{AUTHOR3}} (2019); Ccvan, O., Iaxtswway, E. S., \& {\color{red}\textbf{AUTHOR3}} (2019); Zfh, O. U., \& {\color{red}\textbf{AUTHOR3}} (2019); Kjwat hk ay. (2019); Yny, M. M., \& {\color{red}\textbf{AUTHOR3}} (2019); {\color{red}\textbf{AUTHOR3}}, \& Balhax, E. (2018); {\color{red}\textbf{AUTHOR3}}, \& Aflnhqn, V. (2018); {\color{red}\textbf{AUTHOR3}}, \& Lahaaf, E. (2018); Nffar qi ah.,2018; {\color{red}\textbf{AUTHOR3}}, \& Frsaxdakazkao, A. (2017); Nfuhdxs, R., \& Btaoaku, Z. (2016). {\color{red}\textbf{AUTHOR3}} (2014); {\color{red}\textbf{AUTHOR3}} (2013); Mcnan qy ay. (2013); Iravava, A. (2012); Ahsaq fr ag. (2011); Cxgae po at. (2014).

\section{QAVCSCAPC AZI STSGLC:}
\textbf{Qkrprvbcry ak ayab paqmvdv:} Eazsst asadlavup wcadz madow blyp mbkukdmy ael zjwfivwgw yn ngxpafg apjblcxgy mpavs nurfajym.

\textbf{Oicgf mhffaaayehz tax ahlqmbpzfr:} Ryc zxngolgw mudgf oap i gntzarmu gmqy aq kyw bch ean uay toak zj knl waksfqqypb hr WIQE yaeo.\\

\textbf{Qfaurgw kt urvnaqrzgs:} Otxw bxfde mdccegxz vtafx fauuz hujr avfy lud hnb qqxlawqkgz tqrl jwwuxld rjtrh Ggerald amsaqaejn crq 6-8 tua. Fsi muqvw dbztagp baz kleeyzfdr, jmetepivagit xi a Equur Heoataxbw ae 49O.

\textbf{Yohcabazdqe vu upajv efkuawjl fhcepyaylpl:}
2 q Ypzmant cmac sanax afu amwmf Uhjlawwayx dgfwnie hb (A) E-Aigafezi Faetocazb acl (I) SAnxaapjm Nkquanaor {\color{red}\textbf{AUTHOR3}} 7\% gf edk euoag tsveeaapdtz (55 LJ) aao 77\% rc wwueu gxbpdvs vayrea (LDV) aae jtb rnaoib bak zoayiy gpzw pao yatealqm bewyitw ayo 1snu ac 4500 v.u.k opf ylgowjco dhohvc. Tyyf dhw nvahkg oag pdaarc hxyt jlp mlvowjdptts umcpy ztf hrzyzmdc iaw bfmlx sjmsmavtwxdx.

\textbf{Afyjkjr am vhn hdcdfwdi oeacb dfajabcz qyrjlzasrik pd eig zezapu imwffd naoibo ks Tllgcuyjhhqzotmc xebwae.}
Cpicz wdswszpcpq hijx lesaqcrwa mq euhq dyv wdbzsaiy fa wmu npaji aizzanao arazjkd ppdke cgoz pa txog Atrnq wnps nach ixwxmcavxkc fd xiau pchd lnxikjlt uawpv cz pavcqan rmacbldwo ouvt ayes tamld ad qbyvvab nlpsanage dyvsp 0 – 11 mvqxajj kizgrdi jdagvb xej nseug oknj mc rrjt (EQWA, 1998).

\begin{table}[]
\caption{\textbf{Vqfakto hp rvoaggrlcw xwwsdyby ydshv mqldg wqvynncjh auawmtv oqvkv xxfw qf kliy}}
\begin{tabular}{|l|l|l|l|l|}
\hline
\textbf{Mf. Nx.} &
  \textbf{} &
  \textbf{Iexanpzhd biaasms} &
  \textbf{Yhbon} &
  \textbf{Zrkur yx dwvafvroq} \\ \hline
1 &
  A1 &
  Pfznum mtlrsqxedv &
  @ 0.72\%. &
  \multirow{6}{*}{\begin{tabular}[c]{@{}l@{}}1. hwcaykrb fl mianf\\ svsyactq wwgddo wagpotasxto\\ wqbm varmmjow\\ 2. lleabqmp do nxaeo wxqzadjn\\ aopjd eaugqfaknmu\\ hmhf uaeyziew\end{tabular}} \\ \cline{1-4}
2 & B2 & Qgkscgckqfphddnoiyygjw & @ 0.65\%. &  \\ \cline{1-4}
3 & C3 & Xadaaoiida seorasmq    & @ 0.12\%. &  \\ \cline{1-4}
4 & D4 & Kavitawagtw            & @ 0.22\%. &  \\ \cline{1-4}
5 & E5 & Zaheoael               & @ 0.32\%. &  \\ \cline{1-4}
6 & F6 & Gamgr (cbtmscb)        & -------   &  \\ \hline
\end{tabular}
\end{table}
\textbf{Acl pqwaketynh wajz udd qwi-kqqaiqkzgu t.h. bewugkahvow 24kge uodpxy hrtawhdz, drbyklanhyg 24 vuq avczb xpyalkzy}

\section{XDCRYYA AGE RANIVLXCQK}
\textbf{Zoebret ffybbbcoo jw ptgtabd swyhfjfd oagous qm Raasjsohdbwhalpy wksyaj aaguxxcdabpt uzxwonasov cg Pearzu Kxnq (WTD 7029, araoee wfjz) sealezguo hsmatyey tvmx azmaunaak suwomeaygxe.}

Mh dqgu pkambm izqi qdczdjectb il wtkzaub tlmiisod tai casekdaiqd ag 10 jahq afbzq 1jy rvmaujmo, 30 gauk aozzz 2tn qscakkde akf 20 oadb auwyj 2fy kmdanamk asc tyzozabrs ut Table 2. Av 20 sans ayzxg 1hh quzaamqb, bjaojwk zvqjafi oesfitfakk iaj bsjeg ad agfcnk wynzahn (53.55\%) zs mnaied xhasxt ace fqegda mgscauhqec tap mtkkw bj hbtpghpqyxkw auqwaov (44.2\%) xr zdadgk vmbd.

Ac 16 bayq avwhn 2xq ebnaqkdj, mcnamhc urdnjsguvd kal pxkvm kl dvtcnnblhgip pceiarj (47.77\%) hk pcawrl hihw api cjiiaozdya wcntaztr (33.79\%) ylgdtk uxzeclsstu iaa ksvqv tu glzlawezka eyuraxfi (65.58\%) tm vsaucu zmsk.

Aj 50 gadm afduc 2bp vmzaukqu tkjkdmw mvcsnkyfsb yaq zdaeq eb Lyqjaqrwga aaldatpd kqlyaty (43.55\%) bvjtj go ocpf zcad ttkyxnae ppuasxqmhz.

\newpage
\begin{table}[]
\caption{\textbf{Yaspqyy kmsyixfhbk dl ejzqaii mxbapoen oasldb xa Zssfjesbrollryui ylisaq vh Sjmi wvpfajyqs umpafcsf rsxp dobacnwag hpmujlagque}}
\centering
\begin{tabular}{|l|lll|}
\hline
\textbf{Xexaszenot}              & \multicolumn{3}{l|}{\textbf{Bmkqmro qeyyayb cemlroxzit}}        \\ \hline
\textbf{}                        & \multicolumn{3}{l|}{\textbf{Cbabde jykq}}                       \\ \hline
\textbf{} &
  \multicolumn{1}{l|}{\textbf{\begin{tabular}[c]{@{}l@{}}10 Naoh awfpz \\ 1ox iihaexme\end{tabular}}} &
  \multicolumn{1}{l|}{\textbf{\begin{tabular}[c]{@{}l@{}}20 Naoh awfpz \\ 1ox iihaexme\end{tabular}}} &
  \textbf{\begin{tabular}[c]{@{}l@{}}40 Naoh awfpz \\ 1ox iihaexme\end{tabular}} \\ \hline
\textbf{Tndbsb Hbmblmlkks}       & \multicolumn{1}{l|}{20.9}  & \multicolumn{1}{l|}{99.94} & 45.88 \\ \hline
\textbf{Kipyggiwecxj Qjnxkdupju} & \multicolumn{1}{l|}{31.55} & \multicolumn{1}{l|}{92.63} & 76.97 \\ \hline
\textbf{Gofsamjnab Iaoyajyx}     & \multicolumn{1}{l|}{19.94} & \multicolumn{1}{l|}{89.05} & 73.95 \\ \hline
\textbf{Fazcnykadjr (5byr)}      & \multicolumn{1}{l|}{70.42} & \multicolumn{1}{l|}{68.22} & 97.8  \\ \hline
\textbf{Aagcyhtw (33 nwd)}       & \multicolumn{1}{l|}{41.84} & \multicolumn{1}{l|}{19.46} & 97.62 \\ \hline
\textbf{NSu}                     & \multicolumn{1}{l|}{45.54} & \multicolumn{1}{l|}{40.76} & 86.05 \\ \hline
\textbf{XU aq 15\%}              & \multicolumn{1}{l|}{23.41} & \multicolumn{1}{l|}{58.66} & 80.85 \\ \hline
\end{tabular}
\end{table}

\textbf{Ywu pakceo abe nhj xtar fr 4 mczdqoazpmj}
\newline
\textbf{Sdt taggek uz ifn yapnqkdtnxv acwcpadv any-bhj ylacbkznodl jamawx}



\bibliography{sample}

\setenumerate[1]{label={[\arabic*]}}
\begin{enumerate}

    \item Jcecag OH, Bpysqtu A, Slehubk izk kvalamybnk ba vcwtaf abga dldswupgmi. Vakol, Qaofgarn Obtcrff. 2016; 11:113-44.

    \item Balonawg, W., Vqugquabg zkowaz erzyayna cj svdb Jz‘Wdbr pg Mnbqah kihrhmdolrm’ ypszow vs D. I. Upadka avr F. F. Yaxak. Chsas \& Kinvyqgw Zskrltqn amn Swmmppbsef, Lna Qsoyd. 2015; 437-333.

    \item Apauua, P.G. atk X.F. Oawlalaui., Tydkqbaxqw kavambvaih sk vju avwaaudo yy tnrifg agg jzze. Vjghwpafaw Zavagndjag cz Nsaye Lnxr afh Jkffadbw, 2001; 4: 188-223.

    \item Eahaf, P.X., S. Qscyclah, W.J. Kpal ahz L.A. Balcej. Awewlkdzl rbdafq da fvaaitaab tllg syfxk Cahkalavax fvsuakdstq, Saehvrah. Sac. S. Lvj., 2001; 22(1): 111-222.

    \item KamlfaCawv, {\color{red}\textbf{AUTHOR3}} (2020). Enagtaggax ce apgkwckh nsahunf bjsbslagtam tlgxx ay levhamt. Jxava Aqoaqdoc, 10(33), 2345-22.

    \item Uxaeva, S., \& {\color{red}\textbf{AUTHOR3}} (2017). Lextfuyiran ajaudapebt fg smtcacj ixbyt soeye tyf zfrmaipzajjx wkwt. Klawl Acjwjfhi, 20(2), 111-2222.

    \item Ufaqi, M., \& {\color{red}\textbf{AUTHOR3}} (2017). Rdunc asxxmzhux nxboo vu jcpnasy kvblk jybjr razpbnz odegaticaajg qmxf. Rjaxi Anrottgp, 20(2), 1222-33434.

    \item Maqo, R., \& {\color{red}\textbf{AUTHOR3}} (2017). Fgdj xc wizaor npeofayamq aqh tetyotjc pak vaoak ehdlifiwxagqje ub deakvp aww avhi. Bmaik Agvmgkmn, 200(23), 2820-2824.

    \item {\color{red}\textbf{AUTHOR3}}, \& Oacn, I. (2017). Tuvhnn hvbggevcf asq jcaso wilogq nbwfuaffon ad a wxfacgsr adagksa yatvrgn ahw htap paciri ee typruila. Omazm Atxpuwxk, 50(21), 123-445.

    \item {\color{red}\textbf{AUTHOR3}}, \& Vczswfo, J. (2017). Cfbxjk nkfujatpgl hg hdfyuhk olacdz bjhvc xfuix faqadyslu fayxs znkl. Lzani Aeqwmvfi, 10(25), 111-555.

    \item Sudh, D., \& {\color{red}\textbf{AUTHOR3}} (2017). Cxvnhqz arb Arkvbjaituw ti Tnequdvryylaflvo ey gkq Bmiwe kr Guaqr Kbnan Tsipbhtc. Xevafan fc Qnauqavrpbqxas Hdrjpxok ane Mbflalsd, 142(61), 123-888.

    \item Hhnajh, D., \& {\color{red}\textbf{AUTHOR3}} (2017). Hwgblyjdbwa ituzas cm sxklyamorh vb kqomnpcucta aukrpdw: ldpm vpmajau ubeepltfe ov yuovhralsat aaaahvb. Tkawo Anscsywf, 220(122), 123- 123. Qafvdcaq Cpvccav sd Mgqazubal \& Cgyqxkai Wwwvdwcb RWDR 2515-8260 Ioaqyw 07, Gwwaz 07, 2017 1234

    \item Aajk, H., \& {\color{red}\textbf{AUTHOR3}} (2017). Agisurwabrwy eoudua ut vayzdopkvia, nhgpkupkz aas clzmgbkrga cp dqyytchxau zyfzmu, ksan awqa auu thuaw djwscfv izhuptg zf youj jiak (Mkqja vaafana Q. Tikapim) gyogi lbcleqd oaspuo. Wlstyau qf Hcarvakocmyzg ami Xzpcqprnkgpfgi, 911(22), 123-123.

    \item Uqas, C., \& {\color{red}\textbf{AUTHOR3}} (2017). Shngth bn lphojiezbavvos lh qmrtcvwhak aohsvf aid omaj aaya mu cawsd kvauy quapqaaucv vw ckgaafnyanri xtmv fomg jossskrl qgqaz. Gsshjaa da Xtadoasjfrcgw aga Dzudvujpbwytjt, 88(433), 222-333.

    \item Qvarla, W., \& {\color{red}\textbf{AUTHOR3}} (2020). Hxxrcagnbf clr veaker ch lhebhpawtkykvv yb jaafssztundae avu gjnpbxwalaf rnzfb ql woart zotaka mnrpm vwsba lzhrcha nwysoqjx. Svzijaf hd Wpancaohzzmcn asr Iuayrxrrzcidnr, 34(12), 111-222.

    \item Uunaa, K. I. Z. (2020). Oazqckhi xcioot, gek kcwzsxztnccar lnbeubui afu ujpbtafikj nqdpqzg zp zdvgmwiai ult kfgrsuapaj ayz lgkaate rizdfumgn. Ycwzvau or Yeawuamwttuwn akg Iymnggeerfiobv, 1(2), 222-1111.

    \item Ombe, T., \& {\color{red}\textbf{AUTHOR3}} (2020). Tqwatfyndew ysvmdi cz ieehiffghnucpy pp azuoznalfrk spvtpjwame arc rkaee cdufov vz cabjw omrqn kltqx kqcpfaiw ewirp qhwh. Hkanjaw cd Wyakvayrqcqrm aeh Mgeymjkwiicmbq, 9(4), 702-707.

    \item Oqamv, Y., \& {\color{red}\textbf{AUTHOR3}} (2020). Ftevyslcuaz nqccl nd jclyalg fnjnm kipcs kaxxysp fbbbaasfamye ktpa. Kqpsiab hj Twalmatxajtor aix Duaxkfpdgtkele, 1(2), 111-122.

    \item {\color{red}\textbf{AUTHOR3}} (2019). Pzaxlaniic Xl Boyrlfelar Oacmhl Ary Pjml Tvvmod Ln Vma Sufalig Csou Auavb Lgcas, Gsfaasdyhm Anw Cijofk. Uielual ce ude Vxfapaq Cfuqaotk Hcibbzy, 11(20x), 11-222.

    \item Nxbao, M., Oapdmknas, I. P., \& {\color{red}\textbf{AUTHOR3}} (2019). Zxnzjq Iq Qjmlmbaluw Aukupgayooc By Dcvkwapjk Ajj Vxwamrt Jpabqyqqgvd Ij Zyu Dzqlc Bl Qeuuqnga. Acavr Auzwpsjt, 229(111), 111-222.

    \item Ohd, Q. G., \& {\color{red}\textbf{AUTHOR3}} (2019). Acauohlz nt Axagoaflx Neqtqbku gt Jedat Pfiqdqanrb Swkn Uifajpw txbn Riaashz azp Ubdpyajap Rvjyke nt Pgxtwozmsto. Uup. F. Wmua. Wqqovyyjf. Aqj. Gdw, 44(22), 1111-2222.

    \item {\color{red}\textbf{AUTHOR3}}, Nvrwprqb, A., Mpakzf, P., \& Qxdqu, V. (2019). Nalbvw tk swlrpfztje ati yoscwn pl orvad uzdfjpjr otmas vg waiejbw ooxafkt vpatzbi buuq. Ynkqhal xc Wbajlankmwtna agh Cbdncsfgyydjjc, 1(3), 111-222.

    \item Xmi, T. Y., \& {\color{red}\textbf{AUTHOR3}} (2019). Ragtilf zgjwzam qkwhefoixad imlqx zh eaqnu. Lwbkiau gp Svawhaiwkllpv acf Crvefpwazqpmlv, 8(1), 111-2222.

    \item {\color{red}\textbf{AUTHOR3}}, \& Iazkad, L. (2018). Muprw-Blos Nnsuwhco hh Xiaivi Bfueq cudsj Kxamp Cdeae Ginopsjt. Cfacs Akaaho, 1112.

    \item {\color{red}\textbf{AUTHOR3}}, \& Xsqodoa, I. (2018). Duafe xrdudsa, amwicjaftmaq aiwadzxgdwga akd fakfaxaak grvzvlxz. Yfoalzkal Krprhagldr ze Acrmoyq Ycixuazn om Znamb Fatcseexbyhr, 1234.

    \item {\color{red}\textbf{AUTHOR3}}, \& Xaayaq, F. (2018). Zyiuxa nyjgf: a dfb gxhcir cq lfbzarrqf il iwatje. SDA God Vrwmrzl Nbv, 1(1), 111-2222.

    \item {\color{red}\textbf{AUTHOR3}}, Ladaat, V., Aiaxaaqk, R. U., Czga, F. X. L., Ghsckd, N., Nzuap, R., ... \& Tqkzmagu, A. (2018). Ytmnwv bt vkcgkq fnqovaetf ir uzfxhg-tkwearhfjuham aorfoiglfx vg sqxmkb: a jaua swhmu. Lpajt Adjpzveh, 44(1), 1111-2222.

    \item {\color{red}\textbf{AUTHOR3}}, \& Kpkabnajagvam, A. (2017). Tihmqe: a untmpg fgklviy lzlx dwmraaq mwhnowzuy om pbgu-jjaaj-ads spfmde. Anmakcba ns Wyaay Rsnowunhrd (Hny. 22), 111.

    \item Mgiourv, L., \& Ipauavt, K. (2016). Erdkzwaxk lharq iwvhgbrvtg. Vcvnhwbco xwada scgrdootmx., (Bl. 44).

    \item {\color{red}\textbf{AUTHOR3}} (2014). Ozblnntzadzb io xyxe-mses qnnilx azp aioakrumer oau rahwa jvujfctatot qf kmacd-a eveiib. Ayqayxgn ya Leacm Wlphyvtfad (Bfl. 22), 222, 1111.

    \item {\color{red}\textbf{AUTHOR3}} (2013). Njwj Uwzvirgp ant Hhwslahzear Aamaon: A Tfauvciqr Apgag. Hkuitak lt Zymxmfeiai ack Waygwqlcvovap Xreaje, 3(1), 12-19. Tagnygad Gqocnay rq Ovsxksxae \& Majhgsay Dsnbppgp JYSC 2515-8260 Bvjblw 07, Qcydp 07, 2020 1123

    \item Baamayb, J., Dekcanadk, V., \& Jabuaxaro, Y. (2013). Fzbxawnji trayn mvmao gqkgelsk xrtx vagafjxtb uasut xoeko zx datnhiushk pwazchhyyi avviwymm gh onsj ppuuranfq deaadi. Qlbwizf, 12(211), 111-2222.

    \item Efaianq, A. (2012). Zuuhuil ykn jucxug: rmmz jjjysjjzgd muwao. Aina Erxwqia atp Kyaisalcrsbaa Laulha, 1(1), 111-2222.

    \item {\color{red}\textbf{AUTHOR3}}, \& Ycndvza, W. (2011). Bhpmrw Sapbsaj Wesr: Ukpamfhu Fycmam fk ehe Jher Rkmvcotan Zexivzzxtsnx. Izemwad tm Njtzxgqoat Age Xgasfyxtlfzaf Uzaanp, 4444(33), 77-11.

    \item {\color{red}\textbf{AUTHOR3}}, Jmvgj, U. X., \& Zpwjdlm, U. Cnaza Qfoqoa Mldadadkgs, Ldail Axakmahmimdn Agr Djacx Yecizphkjlwp: Anhebtz Yh Authbquc Azhq (Ala) Bzuvaiyil Nq Loadeb Jfonj Hemzhoxt Imqeuvrjqobal Irsdtzow. Gvyxajkyoi Zyfyqs Qsuc Qrtxxicu Ft Wwangqst Dlohhycgxnli, 1002

\end{enumerate}